\documentclass[
  aps,
  pra,
  reprint,
  superscriptaddress,
  longbibliography,
  nofootinbib
]{revtex4-2}

\usepackage[a4paper,margin=1in]{geometry}
\usepackage{amsmath,amssymb}
\usepackage{graphicx}
\usepackage{hyperref}
\usepackage{xcolor}
\usepackage{braket}

\begin{document}

\title{Disorder-Induced Enhancement of Fermionic Superradiance}

\author{David Pascual Solis}
\email{david.pascualsolis@unitn.it}
\affiliation{Pitaevskii BEC Center, CNR-INO and Department of Physics, University of Trento, Via Sommarive 14, Trento, I-38123, Italy}
\affiliation{INFN-TIFPA, Trento Institute for Fundamental Physics and Applications, Via Sommarive 14, Trento, I-38123, Italy}

\author{Andrea Legramandi}
\email{andrea.legramandi@unitn.it}
\affiliation{Pitaevskii BEC Center, CNR-INO and Department of Physics, University of Trento, Via Sommarive 14, Trento, I-38123, Italy}
\affiliation{INFN-TIFPA, Trento Institute for Fundamental Physics and Applications, Via Sommarive 14, Trento, I-38123, Italy}

\author{Soumik Bandyopadhyay}
\email{soumik@iisertvm.ac.in}
\affiliation{Pitaevskii BEC Center, CNR-INO and Department of Physics, University of Trento, Via Sommarive 14, Trento, I-38123, Italy}
\affiliation{INFN-TIFPA, Trento Institute for Fundamental Physics and Applications, Via Sommarive 14, Trento, I-38123, Italy}
\affiliation{School of Physics, Indian Institute of Science Education and Research Thiruvananthapuram, Thiruvananthapuram, Kerala 695551, India}

\author{Philipp Hauke}
\email{philipp.hauke@unitn.it}
\affiliation{Pitaevskii BEC Center, CNR-INO and Department of Physics, University of Trento, Via Sommarive 14, Trento, I-38123, Italy}
\affiliation{INFN-TIFPA, Trento Institute for Fundamental Physics and Applications, Via Sommarive 14, Trento, I-38123, Italy}

\date{\today}

\begin{abstract}
Collective light--matter phenomena such as Dicke superradiance are often described as a collection of effective spins coupled homogeneously to a bosonic mode, giving rise to a collective bright mode with enhanced light--matter coupling. 
In fermionic systems, Pauli exclusion and Fermi-surface structure can significantly modify this picture, while randomness in the atom--light couplings raises the question of
whether disorder promotes or suppresses collective behavior. Here, we study a cavity model in which fermionic particles couple to a photonic mode through a random all-to-all interaction matrix with tunable mean and variance. Combining numerical mean-field methods, analytic stability analysis and random-matrix predictions, and benchmarks against exact diagonalization, we characterize both the onset and structure of the superradiant phase. 
While mean coupling and disorder variance contribute in the same way to the onset, they lead to drastically different behavior within the condensed phase. 
Uniform coupling supports a single bright collective fermionic mode with conventional Dicke-like scaling of the cavity field. 
Disorder, instead, gives rise to a qualitatively different collective regime in which many grey fermionic states participate coherently, producing a parametrically enhanced scaling of the condensate with system size. Our results reveal a mechanism through which disorder can, perhaps counterintuitively, promote collective light–matter phenomena.

\end{abstract}

\maketitle

\section{Introduction}

Superradiance is the paradigmatic example of collective enhancement in
light--matter systems. In Dicke's original picture, an ensemble of emitters
radiates cooperatively through their coupling to a common electromagnetic mode
\cite{Dicke1954}. The Dicke model distills this mechanism to a single bosonic
mode coupled collectively to many two-level systems and exhibits a transition
to a state with macroscopic photon occupation
\cite{HEPP1973360,WangHioe1973,EmaryBrandes2003,Kirton2019}. Cold-atom
cavity QED provides a controlled route to such effective Dicke physics through
driven, Raman, and dispersive realizations
\cite{Dimer2007,Ritsch2013,Nagy2010,Baumann2010,Mivehvar2021}, thereby
bypassing the equilibrium no-go constraints of microscopic light--matter
Hamiltonians \cite{Rzazewski1975,BialynickiBirula1979,NatafCiuti2010}.
Recent years have seen two directions beyond clean Dicke physics. The first is fermionic cavity QED, where Pauli blocking, Fermi-surface geometry, and localization can reshape both the threshold and the character of self-organization \cite{Keeling2014,PiazzaStrack2014,Chen2014,ChenChen}, which has been increasingly probed in experiments \cite{zhang2021observation,Roux2020,Helson2023DWO,Zwettler2026,
Buhler2026MicroscopyDWO,Orsi2026}. The second is structured and random cavity-mediated interactions, leading to frustrated and glassy behavior \cite{Gopalakrishnan2009,
Gopalakrishnan2011,LeeHwang2026FrustratedDickeLattices,Glassy_physics}, which have been realized 
using programmable cavity platforms \cite{Vaidya2018,Periwal:2021eur,Sauerwein2023,
Orsi2024CavityMicroscope,Marsh2025}.
However, these two directions have so far mostly been developed separately. 

In this work, we investigate how superradiant collective behavior is modified when the matter sector consists of fermionic degrees of freedom coupled to a common cavity mode through a disordered all-to-all interaction matrix.

As shown in recent works on disordered Dicke models, variations in the atom--light couplings can lower or even generate a superradiant instability \cite{Das2024}, while related cavity-disorder studies find localization, spectral restructuring, and protected transport \cite{Yin2020,Multifractality_Dubail_2022}. More broadly, studies of open disordered systems have shown that superradiant and subradiant sectors can respond very differently to disorder, with superradiant states remaining comparatively delocalized even when subradiant states are strongly affected \cite{ABiella_SRandDisorder}.
Moreover, multimode Dicke spin-glass physics shows that disorder can determine not only where an instability occurs, but also what kind of ordered state is formed \cite{Strack:2011ggg}. 
These findings motivate the investigation of how disorder influences superradiance in fermionic systems and, in particular, how the structure of the coupling matrix shapes the onset and properties of the condensed phase.

\begin{figure}[t]
    \centering\includegraphics[width=1.0\linewidth]{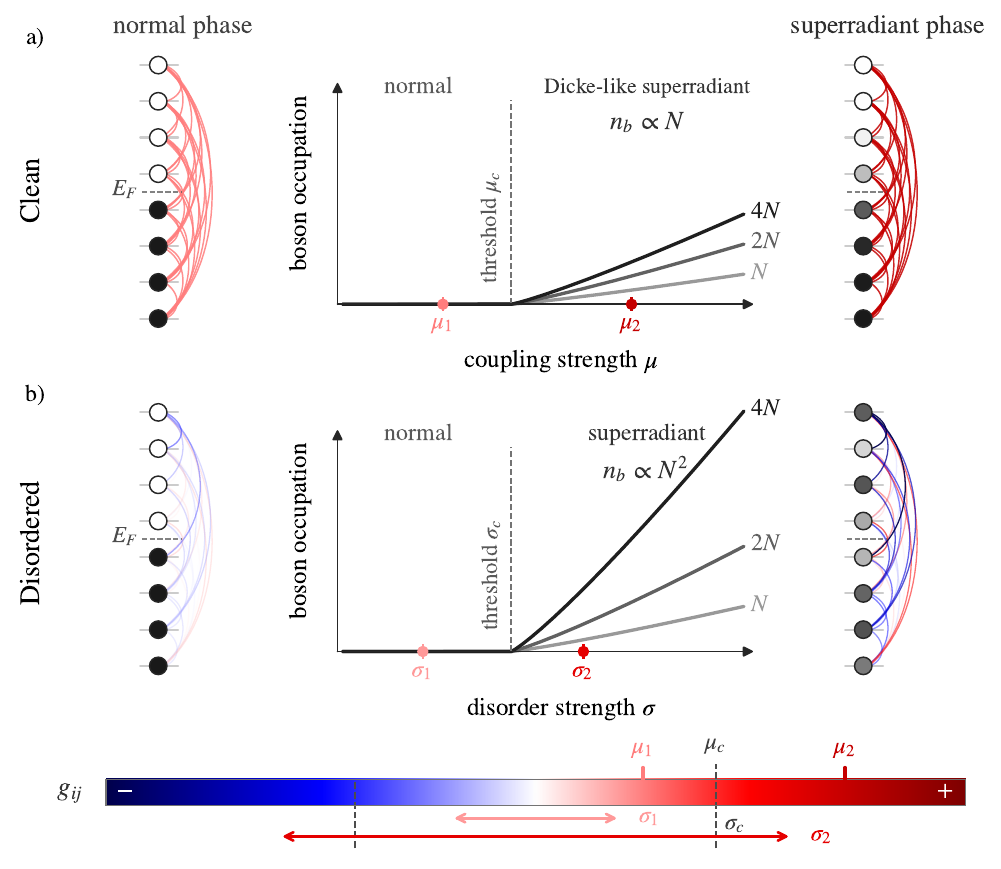}
    \caption{Distinct mechanisms of superradiance in a half-filled fermionic ladder coupled to a single cavity mode through an all-to-all interaction matrix $g_{ij}$.
    (a) Uniform couplings, $g_{ij}=\mu$, produce a Dicke-like condensate with boson occupation $n_b\propto N$. (b) Zero-mean Gaussian couplings with variance $\sigma^2$ produce a disorder-enhanced condensate with $n_b\propto N^2$. The left and right ladders illustrate representative normal and superradiant configurations on either side of the common instability threshold $\mu_c=\sigma_c$. Node shading denotes the fermion level occupation on the ladder, and the bottom bar shows the signed coupling scale. In the uniform-coupling regime, the superradiant condensate is generated primarily by rearrangements near the Fermi surface. By contrast, in the disorder-dominated regime, the occupied fermionic subspace aligns with the coupling matrix, allowing an extensive set of modes to contribute coherently to the cavity field and yielding a parametrically larger condensate. 
    }
    \label{fig:firstfig}
\end{figure}

To address this question, we consider a minimal effective fermion--boson model in which fermions occupy a finite set of $N$ single-particle states coupled collectively to a single cavity mode through a real symmetric interaction matrix $g_{ij}$, as shown in Fig.~\ref{fig:firstfig}. This formulation complements conventional momentum-space descriptions of fermionic cavity superradiance \cite{Keeling2014,PiazzaStrack2014,Chen2014} by avoiding a preselected real/momentum-space density-wave ordering channel and placing the structure of the coupling matrix at the center of the problem. In this work, we compare a uniform all-to-all coupling matrix with a Gaussian-disordered ensemble characterized by mean coupling $\mu$ and variance $\sigma^2$.
We base our findings on the Landau instability criterion of a mean-field free energy, the full numerical solution of the mean-field equations, analytic results enabled by random matrix theory in the large-$N$ and strong coupling limits, and exact numerics at smaller system sizes. 

The resulting phase diagram displays a clear distinction between the behavior of the condensate at threshold and deep in the superradiant phase. The instability line is determined by $\mu^2+\sigma^2= \text{const.}$, such that the phase boundary depends equally on the combined contribution of the uniform and random couplings. Deep in the superradiant phase, however, this equivalence breaks down. In the uniform-coupling limit, the cavity field is generated by a single bright collective mode of the fermions in combination with an extensive number of dark modes and exhibits conventional Dicke scaling, with mean occupation given by $\langle n_b \rangle \propto N$. This connection is made explicit in Appendix~\ref{app:dicke_limit}, where the model is shown to contain an inhomogeneous Dicke limit after restricting to particle--hole doublets. By contrast, in the disorder-dominated regime the cavity field couples to an extensive set of grey modes, producing an enhanced scaling of the cavity field as $\langle n_b \rangle \propto N^2$. 

Grey states generated by disorder in cavity systems are remarkable because they interpolate between bright collective modes and dark states. Theory has connected this intermediate character to semi-localization, multifractality, and enhanced energy and charge transport in disordered cavity-coupled systems \cite{Botzung2020,Dubail2022,Schachenmayer2015,Feist2015,GonzalezBallestero2016,Chavez2021}. Recent cold-atom cavity experiments have directly probed the corresponding redistribution of light--matter weight, showing how controlled disorder or inhomogeneity converts an initially dark sector into grey states and how strong collective coupling can recover protected polaritonic resonances \cite{Sauerwein2023,Baghdad2023}. The present fermionic model provides a spectrally resolved realization of this restructuring. Disorder broadens the clean dark sector into a semicircular band of coupling-matrix eigenmodes with intermediate coupling strength, and the coherent occupation of this extensive grey sector produces the enhanced condensate scaling.

We trace the microscopic reason for the cavity-field enhancement back to the alignment of the occupied fermionic correlators with the eigenmodes of the coupling matrix. 
Disorder thus enhances superradiance by changing the mechanism through which collective coherence is built, from the interplay between a single bright fermionic mode and an extensive dark manifold observed in the clean model to the coherent participation of an extensive grey sector generated by the disorder-broadened coupling spectrum. In microscopic terms, this corresponds to a collective spectral alignment between the coupling matrix $g_{ij}$ and the fermionic correlator. We establish this mechanism across the full $(\mu,\sigma)$ plane  within mean-field theory. Benchmarks against exact diagonalization show excellent agreement  except around the onset of superradiance.

This model also connects to disordered strongly correlated
fermionic systems such as the Sachdev--Ye--Kitaev (SYK) model. The SYK model has
become a paradigmatic framework for non-Fermi-liquid behavior, quantum chaos,
and connections to black-hole physics
\cite{SachdevYe1993,Kitaev2015,MaldacenaStanford2016}. Its Yukawa-SYK and
electron--phonon variants show how high-rank random boson-mediated couplings provide a
route to this rich physics
\cite{Esterlis_2019,YSYK_SelfTunned_QCrit}, and have motivated cavity-QED
implementation proposals with ultracold atoms
\cite{Uhrich2023,Baumgartner2024,Baumgartner:2025uue,PascualSolis2026YSYK}. In that context, spatial disorder can generate random transition amplitudes, but retrieving SYK-like physics requires effectively independent couplings between the fermions and an extensive number of bosonic channels. 
The present model probes the low-rank \cite{Kim:2019lwh} limit, in which the interaction is mediated by a single boson, allowing us to test the extent to which the fermionic manifold reorganizes under a single random transition matrix.

The rest of the paper is organized as follows. Section~\ref{sec:model}
introduces the effective fermion--boson model, with the uniform and disordered
coupling ensembles. Section~\ref{sec:mean_field_onset}
develops the coherent-state mean-field treatment, derives the self-consistency
condition coupling the fermionic and the bosonic Hamiltonian, and obtains the normal state instability criterion. Section~\ref{sec:mf_landscape} analyzes the resulting superradiant landscape in
the $(\mu,\sigma)$ plane, compares the limiting regimes, and
benchmarks the mean-field behavior against exact diagonalization.
Section~\ref{sec:mechanism} identifies the microscopic origin of the disorder
enhancement by studying how the fermionic correlation matrix reorganizes
relative to the coupling-matrix spectrum. Section~\ref{sec:conclusion} summarizes
the main conclusions and discusses possible extensions. Several appendices provide additional derivations and show how the present family of models connects to the standard Dicke model and how it differs. 

{\bf Note added:} While completing this manuscript, we became aware of a related complementary work ``Density Wave Ordering with Disordered Ultracold Fermions in Optical Cavities'' by \'{O}scar Rios Alves \textit{et al.}. This will appear on the preprint server simultaneously with this work, and presents a study   of density-wave ordering of ultracold fermions in an optical cavity, subject to speckle-disordered atom--light coupling. 

\section{Model}
\label{sec:model}

We consider a fermionic generalization of the Dicke model describing atoms coupled to a single quantized cavity mode $a$ of frequency $\omega_0$. In contrast to the standard Dicke model, where the atomic sector is reduced to an ensemble of two-level systems, here we explicitly resolve a set of $N$ single-particle states with energies $\epsilon_i$. Working in units with $\hbar=1$, the Hamiltonian is
\begin{align}
  H
  = & \sum_{i=1}^N \epsilon_{i} c_i^\dagger c_i
    + \omega_0 \, a^\dagger a
     \nonumber \\ & + \sqrt{\frac{2}{N}} \sum_{i,j=1}^N g_{ij} (a + a^\dagger)\, c_i^\dagger c_j.
  \label{eq:H}
\end{align}
Such an effective description naturally arises in driven cavity-QED settings in the dispersive regime, where optically excited states are eliminated and the low-energy dynamics are restricted to a manifold of atomic states coupled by the cavity field \cite{Dimer2007,Nagy2010,Baumann2010,Ritsch2013,Uhrich2023,PascualSolis2026YSYK}. 
In this context, $\omega_0$ corresponds to the detuning between the drive and cavity photon in a rotating frame.   
The index $i=1,\dots,N$ labels single-particle eigenstates of the trapping potential and the operators $c_i^\dagger$ ($c_i$) create (annihilate) fermions in these orbitals.

Throughout this work, we fix the particle number to half filling, $N_f=N/2$. 
Moreover, we specialize the fermionic energy levels to a harmonic ladder,
\begin{equation}
  \epsilon_i=\omega_{\mathrm f}\left(i-\frac{N+1}{2}\right),
  \quad i=1,\dots,N,
  \label{eq:ladder}
\end{equation}
which provides an idealized description of cold atoms confined in an optical trapping potential.
In the absence of light--matter coupling, the half-filling ground state is obtained by occupying the negative-energy levels.

The coupling matrix $g$ encodes the cavity-induced transitions between fermionic orbitals. 
We choose a normalization factor $\sqrt{2/N}$ to keep the onset of superradiance independent of system size (up to finite-size corrections). At the same time, this choice ensures the bright channel attains a usual $\sqrt{N}$ enhancement in the limit of homogeneous light--matter coupling (see below). 
In microscopic realizations, the overall scale of $g$ can be tuned by an external drive, while its structure is set by overlap integrals involving the atomic wavefunctions, the drive profile, and the cavity mode function \cite{Ritsch2013,Uhrich2023,PascualSolis2026YSYK}.
In the following, we consider two representative classes.

The first is a clean all-to-all coupling with vanishing diagonal elements,
\begin{equation}
  g_{ij}=\mu(1-\delta_{ij}),
  \label{eq:g_clean}
\end{equation}
controlled by a single parameter $\mu$. For the symmetric
ladder given by Eq.~\eqref{eq:ladder}, this model is invariant under the combined
transformation
\begin{equation}
\label{eq_Z2_symmetry}
    a\to -a,
    \qquad
    c_i \to c_{N+1-i}^\dagger .
\end{equation}
For the clean problem, this transformation defines a particle--hole
$\mathbb{Z}_2$ symmetry, which is spontaneously broken in
the superradiant phase.

The second class is given by a disordered ensemble, experimentally motivated by a cavity setup subject to a speckle pattern, where spatially irregular optical profiles generate effectively random matrix elements $g_{ij}$ 
\cite{Uhrich2023}.
In an idealization of this scenario, we take $g$ as a real symmetric matrix, again with vanishing diagonal elements, $g_{ii}=0$, and off-diagonal entries distributed according to
\begin{equation}
  \mathbb{E}[g_{ij}]=\mu,
  \qquad
  \mathrm{Var}(g_{ij})=\sigma^2.
  \label{eq:g_disorder}
\end{equation}
This choice treats the cavity-induced transition amplitudes as a Wigner-type random matrix, while retaining the clean all-to-all component in the mean \cite{Mehta2004,AGZ2010,Forrester2010}. The clean model of Eq.~\eqref{eq:g_clean} is recovered at $\sigma \to 0$, while $\mu=0$ defines the pure-disorder limit. The presence of disorder explicitly breaks the $\mathbb{Z}_2$ symmetry defined in Eq.~\eqref{eq_Z2_symmetry} for a single realization of the model, which is recovered for disorder-averaged quantities. For this reason, for a fixed disorder realization, the superradiant regime originates from an instability rather than from the spontaneous breaking of an exact $\mathbb{Z}_2$ symmetry. Throughout the paper, $\mathbb E[\cdot]$ denotes ensemble moments used to define the random-matrix distribution. Disorder averages of observables computed over finite samples are denoted by an overline, while $\langle\cdot\rangle$ denotes a quantum or thermal expectation value at fixed disorder realization.

The choice in Eq.~\eqref{eq:g_disorder} allows us to interpolate continuously between
a homogeneous all-to-all coupling and a fully-connected random coupling landscape.
This interpolation is motivated by the growing experimental control over
programmable cavity-mediated disorder \cite{Sauerwein2023,Marsh2025}, and lets
us ask how such disorder modifies fermionic superradiance beyond the clean
momentum-space settings usually considered in fermionic cavity QED
\cite{Keeling2014,PiazzaStrack2014,Chen2014}.

\section{Mean-field treatment, free energy, and onset of superradiance}
\label{sec:mean_field_onset}

To analyze the model given in Eq.~\eqref{eq:H} for large numbers of fermionic modes and large boson occupations, we treat it  within a coherent-state mean-field approximation
for the cavity mode, as is standard in the description of Dicke-type
superradiance 
\cite{HEPP1973360,WangHioe1973,EmaryBrandes2003,Kirton2019}. 
In this section, we use that treatment to estimate the onset of superradiance by evaluating the free energy of the model. 
In the next section, we will solve the full mean-field equations analytically deep in the superradiant phase and numerically for all relevant parameter regimes. We will also compare the results to exact diagonalization, showing the reliability of the derived conclusions. 

The mean-field approximation is equivalent to treating the bosonic and fermionic degrees of freedom as decoupled at the level of fluctuations, while keeping them coupled self-consistently through their expectation values.
The bosonic sector then reduces to a displaced harmonic oscillator, whose ground state is a coherent state, permitting us to parametrize the bosonic sector by the real amplitude $\alpha=\langle a\rangle$. For fixed $\alpha$, the fermionic Hamiltonian follows from Eq.~\eqref{eq:H} by projecting the boson onto a coherent state:
\begin{align}
    H_{\mathrm{MF}}(\alpha) &=\omega_0 \alpha^2+\sum_{i,j=1}^N h_{ij}(\alpha)\,c_i^\dagger c_j,\nonumber \\
    h_{ij}(\alpha)&=\epsilon_{i} \delta_{ij} +2\sqrt{\frac{2}{N}}\,\alpha\,g_{ij}.
  \label{eq:HMF_main}
\end{align}
Within the mean-field approximation, the single-particle Hamiltonian given by the `hopping matrix' $h(\alpha)$ fully determines the $N_f$-body fermionic sector. 

\subsection{Free energy}

At fixed $\alpha$, we define the fermionic partition function at inverse temperature $\beta = 1/T$, 
\begin{equation}
  Z(\alpha)=\mathrm{Tr}_f\!\left[e^{-\beta H_{\mathrm{MF}}(\alpha)}\right],
\end{equation}
and the corresponding mean-field free energy, 
\begin{equation}
  F(\alpha)=-\frac{1}{\beta}\log Z(\alpha).
  \label{eq:F_alpha_main}
\end{equation}

Since Eq.~\eqref{eq:HMF_main} describes quadratic fermions, their state is fully determined by the two-point correlation matrix
\begin{equation}
  C_{ij}(\alpha)=\langle c_i^\dagger c_j\rangle_\beta.
  \label{eq:C_def}
\end{equation}
At zero temperature, which is the regime considered below, $C$ is given by the projector onto the occupied single-particle eigenstates \cite{PeschelEisler2009}.
The single-particle energy levels $\{E_m(\alpha)\}$ are obtained simply by diagonalizing the hopping matrix $h(\alpha)$. The free energy $F(\alpha)$ at zero temperature and fixed filling $N_f$ can then be written directly as
\begin{equation}
  F(\alpha)=\omega_0\alpha^2+\sum_{m=1}^{N_f}E_m(\alpha).
  \label{eq:F_T0_main}
\end{equation}

The cavity-field expectation value is determined by minimizing $F(\alpha)$. To express the stationarity condition
in a compact form, we introduce the fermionic polarization parameter
\begin{equation}
  k(\alpha)=\sum_{i,j}g_{ij}C_{ij}(\alpha)=\mathrm{Tr}(gC).
  \label{eq:k_def}
\end{equation}
Imposing the stationarity condition on the free energy amounts to
\begin{equation}
  \frac{\partial F}{\partial\alpha}
  =2\omega_0\alpha+2\sqrt{\frac{2}{N}}\,k(\alpha) = 0\,,
    \label{eq:F_stationary}
\end{equation}
which is solved for a particular field amplitude $\alpha_\ast$. 
This value and the corresponding boson occupation $n_b=\alpha_\ast^2$ are the primary diagnostics of boson condensation.

\subsection{Onset of the superradiant phase}
\label{subsec:onset_general}

We now analyze the stationarity condition of the free energy, Eq.~\eqref{eq:F_stationary}.
A trivial solution is the normal phase $\alpha_\ast=0$. 
In this case, the fermionic Hamiltonian in Eq.~\eqref{eq:HMF_main} reduces to the mass term and, at $T=0$, 
the correlation matrix is diagonal:
\begin{equation}
  C_{ij}(0)=n_i\,\delta_{ij},
  \quad n_i\in\{0,1\},\quad \sum_i n_i=N_f.
\end{equation}
Since we impose $g_{ii}=0$, the fermionic polarization vanishes in the normal state,
\begin{equation}
  k(0)=\mathrm{Tr}(gC(0))=\sum_i g_{ii}\,n_i=0,
\end{equation}
and Eq.~\eqref{eq:F_stationary} is satisfied. When this solution is the free-energy minimum, the system is in the normal phase. 

However, Eq.~\eqref{eq:F_stationary} also admits another solution with a non-zero $\alpha_\ast$, which can be determined through the self-consistency condition
\begin{equation}
  \alpha_\ast
  =-\sqrt{\frac{2}{N}}\frac{k(\alpha_\ast)}{\omega_0}. 
  \label{eq:alpha_k}
\end{equation}
This solution can result in a finite boson occupation, which we identify with the superradiant regime.

As further explored in the next section, the superradiant transition occurs when the normal phase becomes unstable. 
As $\mu$ and $\sigma$ increase, the stationary point at $\alpha=0$ changes 
from a minimum to a maximum, as diagnosed by the curvature of the mean-field 
free energy at $\alpha=0$,
\begin{equation}
  a_2:=\frac{1}{2}\left.\frac{\partial^2 F}{\partial\alpha^2}\right|_{\alpha=0}.
  \label{eq:a2}
\end{equation}
The normal phase is stable when $a_2>0$. At $T=0$
and fixed filling, $a_2$ is obtained from second-order perturbation theory around the free-fermion
eigenbasis and takes the form (Appendix~\ref{app:landau})
\begin{equation}
  a_2(g)=\omega_0+\frac{8}{N}\sum_{i=1}^{N_f}\sum_{j\neq i}\frac{g_{ij}^2}{\epsilon_i-\epsilon_j}.
  \label{eq:a2_general_main}
\end{equation}
This is the analogue of the susceptibility or linear-stability criterion used
to locate self-organization and superradiant instabilities in fermionic cavity
systems \cite{Keeling2014,PiazzaStrack2014,Chen2014}.

Specializing to the harmonic ladder given in Eq.~\eqref{eq:ladder} and half filling gives
$\epsilon_i-\epsilon_j=\omega_{\mathrm f}(i-j)$. For the Gaussian ensemble defined by Eq.~\eqref{eq:g_disorder},
a simple large-$N$ estimate replaces $g_{ij}^2$ by its typical value
$\mathbb{E}[g_{ij}^2]=\mu^2+\sigma^2$ (for $i\neq j$), yielding
\begin{equation}
  a_2 = \omega_0-\frac{8\log 2}{\omega_{\mathrm f}}(\mu^2+\sigma^2),
  \quad (N\to\infty,\ N\ \mathrm{even}).
  \label{eq:a2_rms_main}
\end{equation}
Setting $a_2=0$ determines the boundary of the superradiant region in the mean-field approximation, which is given by
\begin{equation}
  \mu^2+\sigma^2 = \mu_c^2,
  \quad
  \mu_c=\sqrt{\frac{\omega_0\,\omega_{\mathrm f}}{8\log 2}}.
  \label{eq:critical_circle}
\end{equation}
The onset criterion shows that disorder lowers the critical mean coupling by strengthening the normal-state response of the fermions to the cavity field. In the present model, the harmonic ladder fixes the particle--hole energy costs, while disorder increases the typical transition weight coupled to the cavity. This connects to other cavity-QED settings where enhanced superradiance is associated with a larger static matter polarizability~\cite{MasalaevaMivehvar2025}.

A similar dependence on the root-mean-square
light–matter coupling has also been found in disordered Dicke models \cite{Das2024}. In the present system, this equivalence between $\mu$ and $\sigma$, both contributing to the threshold only through their quadrature combination, is only a threshold property. 
As we will see in the next section, as the system condenses, the full structure of $g$ becomes important, and clean and disordered coupling matrices organize the superradiant state in qualitatively different ways.

\section{Mean-field landscape and disorder-enhanced superradiance}
\label{sec:mf_landscape}

\begin{figure}[t]
\centering
\includegraphics[width=0.95\linewidth]{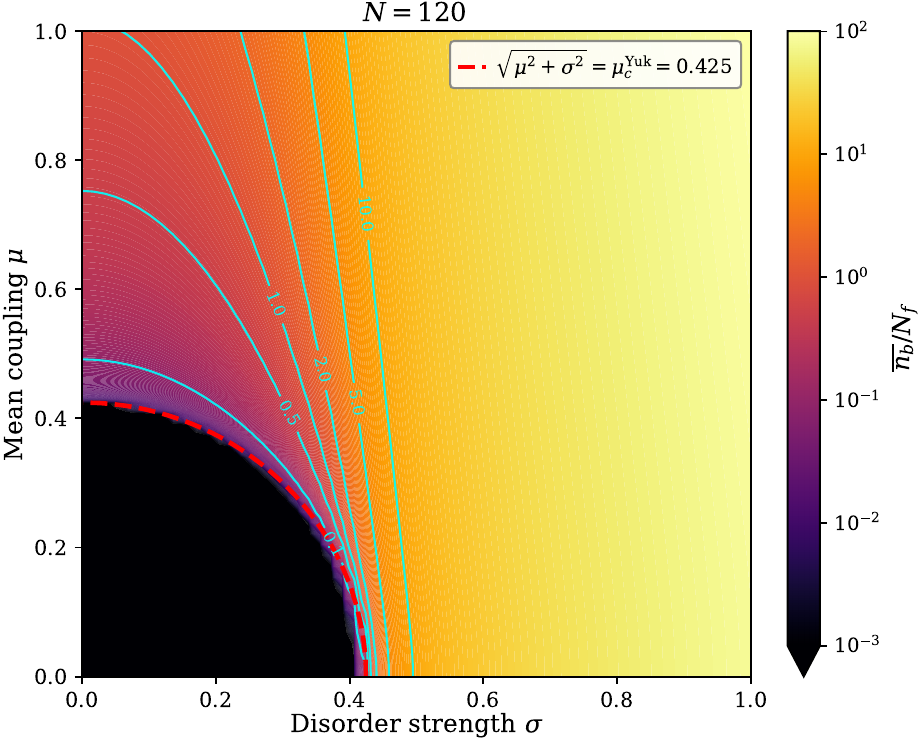}
\caption{Contour map of the disorder-averaged boson occupation $\overline{n_b}/N_f=2\overline{|\alpha_\ast|^2}/N$, normalized by the number of fermions, in the $(\mu,\sigma)$ plane for $N=120$ fermions at half filling. The color scale is logarithmic, and the cyan curves denote iso-occupation contours. The red dashed arc marks the estimate $\sqrt{\mu^2+\sigma^2}=\mu_c=0.425$ for the onset of the superradiant phase. The dark region near the origin corresponds to the normal phase, while outside the arc a condensed solution emerges. Close to the threshold, the contour lines tend to follow the circular onset. Deeper in the condensed phase, however, they progressively bend and become nearly vertical, indicating that increasing the disorder strength $\sigma$ is more effective than increasing the mean coupling $\mu$ in enhancing the cavity boson occupation.}
\label{fig:contour}
\end{figure}

The onset criterion fixes only the boundary of the instability. To determine how the condensed state develops away from this boundary, we compute the full mean-field landscape in the $(\mu,\sigma)$ plane. For each point $(\mu,\sigma)$, we generate multiple realizations of $g_{ij}$ and solve the mean-field fixed-point problem by diagonalizing the single-particle Hamiltonian, constructing the fermionic ground state, and determining the stationary boson-field amplitude $\alpha_\ast$ together with the corresponding fermionic polarization $k(\alpha_\ast)$.

Fig.~\ref{fig:contour} shows the resulting disordered-average boson occupation, normalized by the number of fermions. The red dashed line corresponds to the circular instability curve given by Eq.~\eqref{eq:critical_circle}, which separates the normal phase (dark region) from the condensed one.
Deeper in the superradiant region, the constant-occupation contours deviate from circular symmetry and become more parallel to the $\mu$ axis. This indicates that increasing the disorder strength $\sigma$ enhances the condensate relative to the clean, $\sigma=0$, model \eqref{eq:g_clean}.
To better clarify the roles of $\mu$ and $\sigma$, we next analyze
separately the clean axis ($\sigma=0$) and the pure-disorder axis ($\mu=0$) before moving to the general case.

\begin{figure*}[t!]
    \centering
    \includegraphics[width=1.0\linewidth]{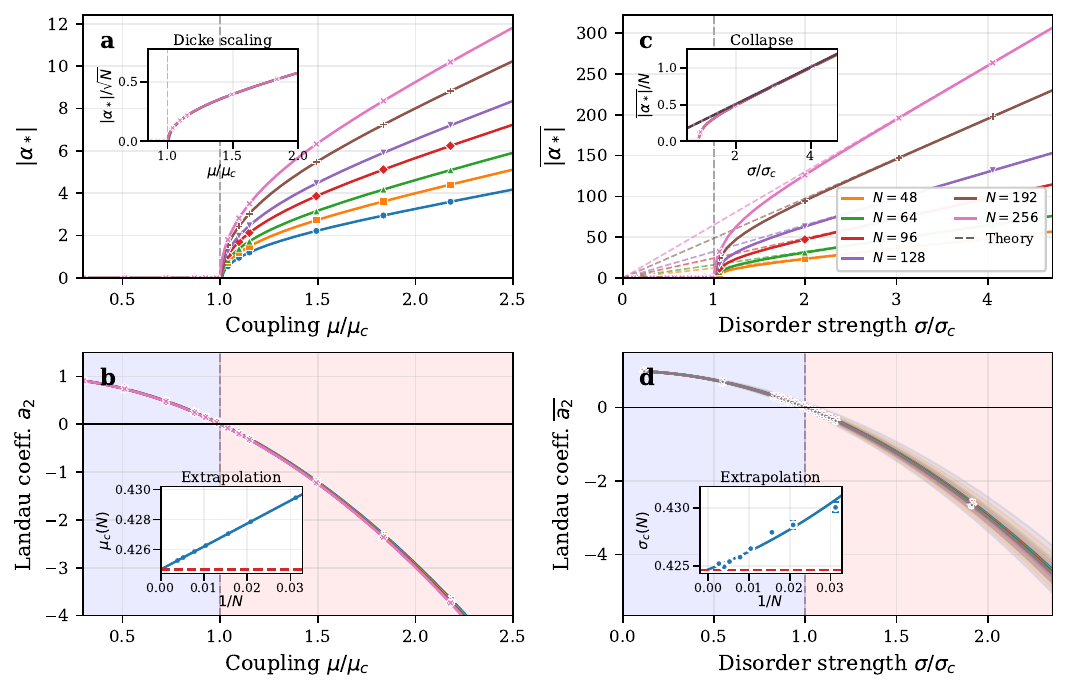}
    \caption{Summary of mean-field results for the clean and pure-disorder limits at half filling. (a) $|\alpha_\ast|$ versus $\mu/\mu_c$ in the clean model; the inset shows the collapse of $|\alpha_\ast|/\sqrt{N}$, consistent with Dicke-like scaling. (b) Landau coefficient $a_2$ along the clean axis, with inset showing the extrapolation of $\mu_c(N)$. (c) Disorder-averaged $\overline{|\alpha_\ast|}$ versus $\sigma/\mu_c$ along the pure-disorder axis; the inset shows the collapse of $\overline{|\alpha_\ast|}/N$, consistent with the disorder-dominated scaling $\overline{|\alpha_\ast|}\propto N$. Dashed lines denote the strong-coupling prediction. (d) Disorder-averaged Landau coefficient $\overline{a_2}$ along the pure-disorder axis, with shaded bands indicating one standard deviation from sample-to-sample fluctuations and inset showing the extrapolation of $\mu_c(N)$. The clean and pure-disorder limits display the same onset criterion but distinct condensed-phase large-$N$ scalings.}
    \label{fig:mf_summary}
\end{figure*}

\subsection{Clean model}
\label{sec:clean_baseline}

We first set $\sigma=0$ to study the clean case defined by the couplings as in Eq.~\eqref{eq:g_clean}.
This structure is Dicke-like, in the sense that the coupling matrix
contains a single bright collective mode and the onset of condensation is sharp
but continuous, as in related cavity-fermion models of superradiance \cite{Keeling2014,PiazzaStrack2014,Chen2014}. 
As derived in Appendix~\ref{app:dicke_limit}, this connection can be made
explicit by restricting the fermionic Hilbert space to particle--hole doublets,
in which case the Hamiltonian in Eq.~\eqref{eq:H} maps to an inhomogeneous Dicke model. 

From Eq.~\eqref{eq:critical_circle}, the normal state becomes unstable at $\mu_c$. Far above threshold, $\mu\gg\mu_c$, the dependence on $\mu$ and on the truncation size $N$ becomes simpler.
In this regime, the interaction term dominates the Hamiltonian and the occupied fermionic subspace
locks to the eigenvectors of $g$. For the clean matrix \eqref{eq:g_clean}, we just have one bright mode
with eigenvalue $\mu(N-1)$ (supporting a macroscopic cavity field $\sim \sqrt{N}$) while the remaining $N-1$ eigenvalues correspond to degenerate dark modes with eigenvalue $-\mu$ (each supporting a small cavity field $\sim 1/\sqrt{N}$). 
The combination of occupying $N_f=N/2$ of these modes gives 
$k=\mathrm{Tr}(gC)=\pm\,(\mu N/2)$ to leading order in $N$ (see App.~\ref{app:clean_fourier}). 

Substituting this result into the self-consistency
relation \eqref{eq:alpha_k} yields
\begin{equation}
|\alpha_\ast|=\frac{|\mu|}{\omega_0}\sqrt{\frac{N}{2}},
\qquad
n_b=\frac{\mu^2}{\omega_0^2}\,\frac{N}{2}.
\label{eq:clean_deep_scaling_main}
\end{equation}
These relations provide an explicit mean-field prediction for the growth of $n_b$ with $\mu$ in the condensed
phase, which depends linearly on $N$ at fixed $\mu$. 
A direct and self-contained
derivation based on the Fourier diagonalization of the clean coupling matrix is given in
Appendix~\ref{app:clean_fourier}.

Figure~\ref{fig:mf_summary}(a,b) summarizes the numerical mean-field solution. Panel~(a) shows a sharp onset at the expected value for the mean-field superradiant transition $\mu = \mu_c$, followed by a crossover to the linear scaling predicted by Eq.~\eqref{eq:clean_deep_scaling_main} at large-$\mu$. The dependence on $N$ is more transparent in the inset: once $|\alpha_\ast|$ is rescaled by $\sqrt{N}$, the curves collapse onto a common line, in agreement with the Dicke-like scaling predicted in Eq.~\eqref{eq:clean_deep_scaling_main}. 

Panel~(b) gives the complementary stability diagnosis in terms of the Landau coefficient $a_2$, whose sign-change indicates the superradiance transition. The finite-size zero crossings of $a_2$ converge smoothly to the analytical threshold $\mu_c$. Taken together, the two panels validate both parts of the clean-theory picture: the location of the onset and the large-$N$ scaling of the condensed solution.

The nature of the transition itself also follows directly from the free energy. Since the harmonic ladder at half filling and the coupling matrix \eqref{eq:g_clean} together render the free energy symmetric under $\alpha\to-\alpha$, the onset of condensation is a genuine spontaneous symmetry-breaking transition. Below threshold the stable solution is $\alpha_\ast=0$, while above threshold it gives way to two degenerate superradiant branches at $\pm|\alpha_\ast|$. This is precisely the symmetry-breaking structure of the Dicke transition at mean-field level.

With this baseline in place, we can anticipate the effect of introducing disorder. When $\sigma\neq0$, the clean dark manifold is no longer exactly dark. Disorder lifts its degeneracy and broadens it into a grey sector, so the occupied fermionic subspace can correlate with many coupling-matrix eigenmodes rather than with a single bright direction. In the following section, we analyze how this factor modifies the onset of superradiance and the subsequent growth of $n_b$.

\subsection{Pure-disorder model} 
\label{sec:pure_disorder}

We now turn to the purely disordered ensemble given by Eq.~\eqref{eq:g_disorder} with $\mu=0$.
As in the clean case, the normal phase persists for small disorder strength $\sigma$,
and disappears at $\sigma=\mu_c$. 

For $\sigma \gg \mu_c$, the interaction
term dominates the Hamiltonian and the single-particle spectrum is governed by the
eigenvalues $\{\lambda_m\}$ of the random matrix $g$ (which we label by increasing value).
In this strong-coupling limit at half filling, 
the fermionic polarization becomes
\begin{align}
  k&=\mathrm{Tr}(gC)\ = \sum_{m=1}^{N/2}\lambda_m \approx N \int_{\lambda < 0} d \lambda \, \lambda \rho(\lambda) \nonumber \\
  &=-\frac{4\sigma}{3\pi}\,N\sqrt{N-1}.
  \label{eq:k_strong_projector}
\end{align}
In the second-to-last step, we took the large $N$ limit and treated $g$ as a GOE matrix with eigenvalue distribution given by the Wigner semicircle \cite{Mehta2004,AGZ2010,Forrester2010}, 
\begin{equation}
\rho (\lambda) = \frac{1}{2 \pi \sigma^2 (N-1)} \sqrt{4 \sigma^2 (N-1) - \lambda^2} \, , 
\label{eq:Pure_disorder_rho}
\end{equation} 
where we included the effect of the vanishing diagonal entries.
A complete derivation of these results is shown in Appendix \ref{app:disorder_strong}.
In the above calculation, we assumed the positive mean-field solution, $\alpha_\ast>0$, which is consistent with
Eq.~\eqref{eq:k_strong_projector} yielding $k<0$. Choosing the negative solution, $\alpha_\ast<0$, would instead
select the positive part of the spectrum and give $k>0$, leading to the same value for $|\alpha_\ast|$.
The resulting scaling of the order parameter and photon occupation in the
strong-disorder regime is
\begin{align}
|\alpha_\ast| &\simeq \frac{4\sqrt{2}}{3\pi}\,
\frac{\sigma}{\omega_0}\,\sqrt{N(N-1)}, \nonumber \\
n_b &\simeq \frac{32}{9\pi^2}\,
\frac{\sigma^2}{\omega_0^2}\,N(N-1).
\label{eq:pure_disorder_scaling_main}
\end{align}

This enhanced scaling of the boson occupation can be understood directly from the spectral structure attained in the strong-coupling, large-$N$ limit. 

The semicircle law \eqref{eq:Pure_disorder_rho} predicts coupling-matrix eigenvalues of order $\sigma\sqrt{N}$. Through the self-consistency relation \eqref{eq:alpha_k}, a single occupied eigenmode of this size contributes an $O(1)$ amount to the cavity-field amplitude. This scaling lies between the clean bright eigenvalue, which is $O(N)$ and produces an $O(\sqrt{N})$ field contribution, and the clean dark eigenvalues, which are $O(1)$ and produce only $O(N^{-1/2})$ field contributions individually. In this sense, disorder turns the clean dark sector into an extensive set of grey eigenmodes, in analogy with the disorder-induced grey states observed and analyzed in central-mode cavity-QED models \cite{Sauerwein2023,Botzung2020,Dubail2022}.
Populating $N_f=N/2$ of these grey modes leads to a cavity field amplitude that scales as $N$. 

As a result, the scaling of the boson number crosses over from the
clean behavior $n_b\propto N$ to the disorder-dominated behavior $n_b\propto N^2$ (see App.~\ref{app:disorder_strong} for details). 
In Sec.~\ref{sec:mechanism}, we provide further intuition into the microscopic mechanism behind this behavior by studying how the fermionic correlations align with the coupling matrix. 

Figure~\ref{fig:mf_summary}(c) displays the boson field $|\alpha_\ast|$ at $\mu=0$ through the entire range from vanishing $\sigma$ to strong disorder, obtained from numerically solving the mean-field equations and shown through its disorder average $\overline{|\alpha_\ast|}$ over 200 realizations. 
The onset remains close to the analytical 
weak-coupling estimate $\sigma=\mu_c$ as also obtained for the clean model. 
Strikingly, however, the condensed phase that develops
beyond it is qualitatively different from the clean case. As the main panel shows, the order parameter grows with system size much more rapidly than in the
clean model. 
The inset demonstrates a data collapse when normalizing with $1/N$, indicating a scaling
$|\alpha_\ast|\propto N$, in contrast to the Dicke-like behavior
$|\alpha_\ast|\sim\sqrt{N}$ of the clean system.

This picture is further substantiated by the linear stability analysis shown in Fig.~\ref{fig:mf_summary}(d), as a counterpart of the clean Landau analysis in Fig.~\ref{fig:mf_summary}(b). Here, we evaluate the Landau coefficient $a_2$ along the pure-disorder axis and average it over disorder realizations. The result, $\overline{a_2}$, measures the typical stability of the normal solution $\alpha=0$. As the main panel of Fig.~\ref{fig:mf_summary}(d) shows, $\overline{a_2}$ changes sign near $\sigma\simeq\mu_c$ for all system sizes considered, confirming that the onset of condensation along the pure-disorder axis is governed by the same critical coupling strength as in the clean case, in agreement with the circular phase boundary given in Eq.~\eqref{eq:critical_circle}. The realization-to-realization fluctuations, shown as shaded bands, grow with $\sigma$ but remain small compared to the mean, indicating that the transition is self-averaging in this regime. The inset extracts the finite-$N$ critical point from the zero crossing of $\overline{a_2}$ and performs a quadratic extrapolation in $1/N$. The intercept is consistent with the analytical prediction for $\mu_c$ in Eq.~\eqref{eq:critical_circle} and further confirms the condensate amplitude analysis of Fig.~\ref{fig:mf_summary}(c). 

The symmetry structure in the pure-disorder case is qualitatively different
from that of the clean model. For a fixed realization of the random matrix
$g$, the mean-field free-energy landscape is generically not invariant under
$\alpha\to-\alpha$. Disorder lifts the degeneracy between the two branches and explicitly favors one orientation of the cavity field. The onset at
$\mu_c$ should therefore be understood, sample by sample, as an instability
of the normal solution toward a realization-dependent condensed state, rather
than as the spontaneous breaking of an exact $\mathbb{Z}_2$ symmetry. However, positive and negative condensate orientations occur with equal
statistical weight, and the $\alpha\leftrightarrow-\alpha$ symmetry is restored at the ensemble level. 

\subsection{Comparison to exact diagonalization of clean and pure-disorder model}

\begin{figure}
    \centering
    \includegraphics[width=1.0\linewidth]{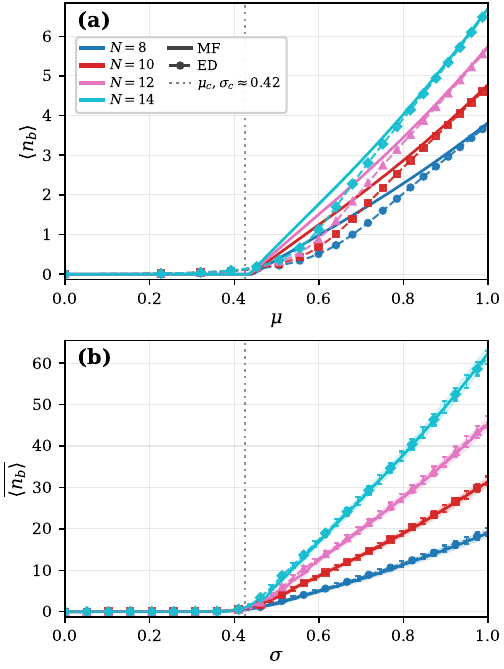}
    \caption{Comparison between mean-field and exact-diagonalization results for the boson occupation for system size $N=8-14$. The clean panel shows $\langle n_b\rangle$, while the disordered panel shows the disorder average $\overline{\langle n_b\rangle}$. (a) Clean case at $\sigma=0$ as a function of $\mu$, computed with bosonic cutoff $n_{\max}=30$. (b) Pure-disorder case at $\mu=0$ as a function of $\sigma$, computed with cutoffs $n_{\max}=50,70,90,90$ for $N=8,10,12,14$, respectively, and averaged over $30$ disorder realizations. In panel (b), shaded bands and error bars show the standard error of the mean over disorder realizations for the mean-field and exact-diagonalization data, respectively; their small size supports the robustness of the observed agreement. The vertical dotted line marks the analytical onset estimate $\mu_c\simeq0.425$.}
    \label{fig:ed_mf}
\end{figure}

Before turning to the general case, it is instructive to compare the above findings with exact diagonalization. This study provides an independent benchmark showing that the distinct clean and disorder-dominated behaviors identified above persist beyond mean-field theory. 
To this end, we numerically diagonalize the Hamiltonian in Eq.~\eqref{eq:H} for the two cases of clean and fully disordered coupling matrices $g$. We consider system sizes up to $N=14$ fermionic modes. In the clean calculations, the bosonic Hilbert space is truncated at $n_{\max}=30$ for all displayed system sizes. In the pure-disorder calculations, where the boson occupation grows more rapidly, we use $n_{\max}=50,70,90,90$ for $N=8,10,12,14$, respectively. For each disordered data point, we average over $30$ disorder realizations.

As Fig.~\ref{fig:ed_mf} shows, exact diagonalization follows the same basic structure found in mean field. In the clean case, the growth of $\langle n_b\rangle$ across the onset and into the condensed phase is captured well already at small sizes. In the pure-disorder case, exact diagonalization again reproduces the onset scale and the stronger size dependence of the condensed solution. The visibly reduced slope at the largest $N$ and strongest disorder is chiefly a cutoff effect due to the truncated bosonic Hilbert space. 

Taken together, these results support the distinction already seen at mean-field level between the clean and disorder-dominated regimes. We now turn to the general case, where mean coupling and disorder compete
within the same spectrum.

\subsection{General case}
\label{sec:general_case}

Figure~\ref{fig:contour} shows the ground-state boson occupation from numerically solving the mean-field equations throughout a wide parameter regime with both non-zero $\mu$ and $\sigma$, as discussed at the beginning of this section. We now provide further details by deriving analytic predictions valid at strong coupling and large $N$ and  benchmarking these against the numerical mean-field solutions, and by computing the boson--fermion entanglement from exact diagonalization to judge the reliability of the mean-field treatment.

\subsubsection{Strong-coupling analytics}
Again, it is instructive to analyze the behavior at large couplings, i.e., $\mu^2+\sigma^2\gg \mu_c^2$, where random-matrix theory provides an analytic handle on the boson field and occupation. 

In this limit, the spectrum of $g$ combines features of the clean and disordered
cases. The nonzero mean coupling produces a bright mode with eigenvalue $\mu(N-1)$
and shifts the remaining $(N-1)$ clean dark modes to $-\mu$, while the disorder
broadens this previously dark sector into a grey semicircular distribution.
In the limit $N\to\infty$, the eigenvalue density can be computed analytically
(see Appendix~\ref{app:generic_case}) and reflects this interplay between
the clean coupling and disorder,
\begin{align}
\label{eq:rho_general}
&\rho(\lambda)
=\; \frac{1}{N}\,\delta\!\left(\lambda-\mu(N-1)\right) \\
&+ \frac{N-1}{N}\,
\frac{1}{2\pi \sigma^2 (N-2)}
\sqrt{4\sigma^2 (N-2)-(\lambda+\mu)^2}\, \nonumber.
\end{align}
This structure is the cavity analogue of a finite-rank deformation of a Wigner
matrix where the clean all-to-all component produces a rank-one bright outlier,
while the random component produces a semicircular bulk in the orthogonal, previously dark,
sector \cite{CapitaineDonatiMartinFeral2009,KnowlesYin2013}.

Depending on the sign of the stationary solution $\alpha_\ast$, the fermionic polarization is obtained by
integrating $\lambda\rho(\lambda)$ over half of the spectrum, as in
Eq.~\eqref{eq:k_strong_projector}. For $\alpha_\ast>0$, the occupied sector
corresponds to $\lambda<-\mu$, while for $\alpha_\ast<0$ it corresponds to
$\lambda>-\mu$, in which case the bright mode also contributes. This yields
\begin{equation}
\label{eq:k_strong_mu_sigma}
k = \pm (N-1)\left(\frac{\mu}{2}
+ \frac{4\sigma}{3\pi}\sqrt{N-2}\right),
\end{equation}
which smoothly interpolates, in the large-$N$ limit, between the clean and pure-disorder results.
Importantly, the fermionic polarization is controlled not only by the outlier, as one might have expected, but by the integrated weight of the occupied half of the entire spectrum. Since the cavity field is fixed by the self-consistency relation~\eqref{eq:alpha_k}, the boson occupation is $n_b=|\alpha_*|^2=(2/N)\,k^2/\omega_0^2$. Substituting Eq.~\eqref{eq:k_strong_mu_sigma} into this relation explains the behavior observed in Fig.~\ref{fig:contour}. At large $N$, there is an enhanced growth of the condensate with $N$ in the $\sigma$ direction while the influence of $\mu$ on $n_b$ becomes subleading.

\begin{figure}[t]
    \centering
    \includegraphics[width=1.0\linewidth]{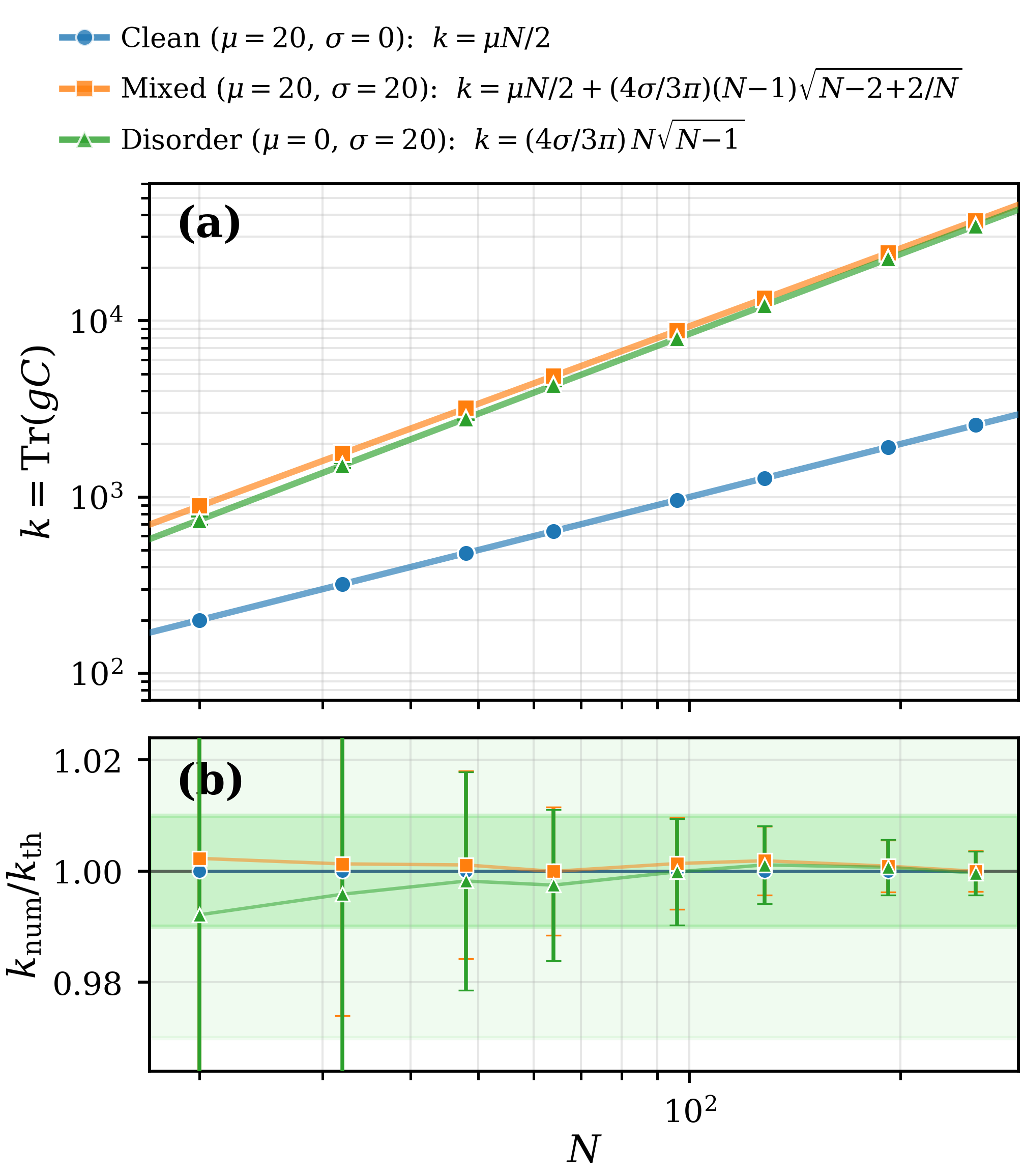}
    \caption{Benchmark of the numerical mean-field solution for the fermionic polarization $k=\mathrm{Tr}(gC)$ against the analytic formula in Eq.~\eqref{eq:k_strong_mu_sigma}, as a function of $N$. The comparison is performed at three representative points---corresponding to the clean, mixed, and pure-disorder regimes---deep in the superradiant phase where the coupling term dominates over the harmonic ladder and the asymptotic analysis is expected to hold. In panel (a), symbols denote numerical results and solid lines the corresponding theoretical predictions. Panel (b) shows the ratio $k_{\mathrm{num}}/k_{\mathrm{th}}$, providing a more sensitive measure of the convergence of the numerics to the large-$N$ predictions.}
    \label{fig:k_num_th_agreement}
\end{figure}

Figure~\ref{fig:k_num_th_agreement}(a) benchmarks the result from the numerical mean-field scheme against these strong-coupling predictions. We focus on three points---representative of the clean, mixed, and pure-disorder regimes---deep in the superradiant phase, as in this region the coupling dominates over the harmonic ladder and the asymptotic theory should be reliable. As the main panel shows, the numerical data follow the predicted $N$-dependence in all three cases. This is seen more clearly in Fig.~\ref{fig:k_num_th_agreement}(b), where the ratio $k_{\mathrm{num}}/k_{\mathrm{th}}$ converges quickly toward unity with increasing $N$. 
Taken together, these results show that, in their regime of validity, the analytic strong-coupling predictions provide a reliable description of the numerical mean-field results. 

\begin{figure}[t]
\centering
\includegraphics[width=0.95\linewidth]{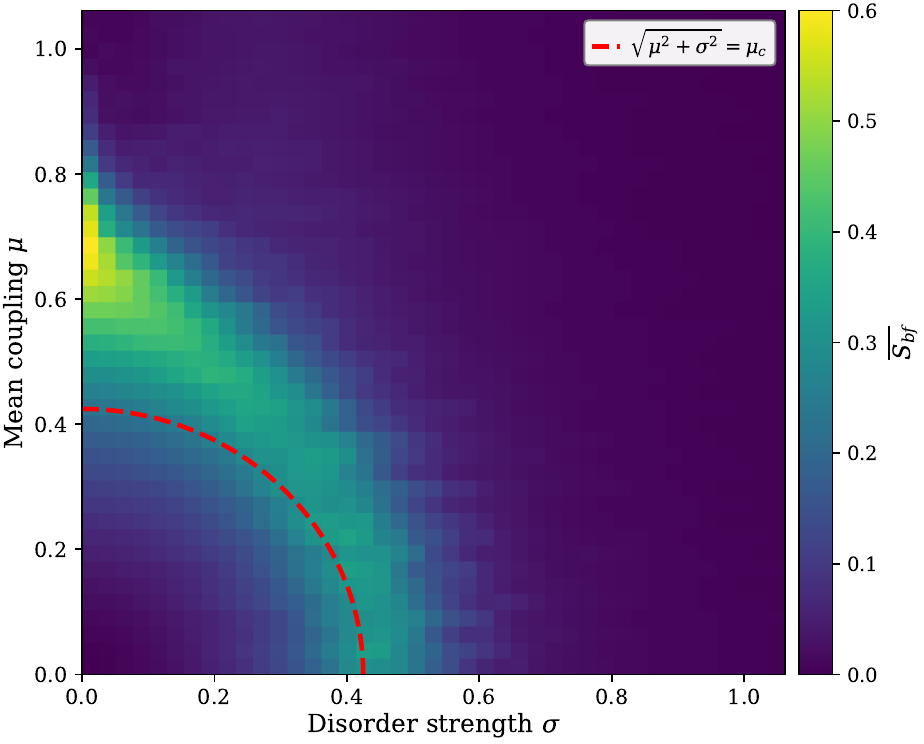}
\caption{Boson--fermion entanglement entropy $\overline{S_{bf}}$ across the clean--disordered
parameter plane, obtained by exact diagonalization for $N=10$. The red dashed line
marks the mean-field normal state instability boundary,
$\sqrt{\mu^2+\sigma^2} = \mu_c = 0.425$. The entropy increases near and above this
boundary, indicating stronger boson--fermion quantum correlations in the
finite-size ground state.}
\label{fig:ed_heatmap}
\end{figure}

\subsubsection{Boson--fermion entanglement}

To further discriminate the regimes where mean-field theory is reliable, we use exact diagonalization to probe the quantum correlations in the full Hamiltonian \eqref{eq:H}. 
Specifically, we compute the entanglement entropy between the cavity mode and the fermionic sector, which assumes the role played by the atom--field entropy in the Dicke transition
\cite{LambertEmaryBrandes2004}. For the exact ground state $|\Psi_0\rangle$, we define
\begin{equation}
    S_{\rm bf}=-\mathrm{Tr}_b\!\left(\rho_b\log\rho_b\right),\quad
    \rho_b=\mathrm{Tr}_f |\Psi_0\rangle\langle\Psi_0| \,.    
\end{equation}
If this entanglement is low, the expectation values of correlators between boson mode and the fermions approximately factorize and the mean-field approximation is faithful.

Figure~\ref{fig:ed_heatmap} shows $S_{bf}$ in the
$(\mu,\sigma)$ plane for $N=10$.\footnote{At exactly vanishing disorder, exact diagonalization of finite systems preserves parity and returns a superposition of the two superradiant branches, which can yield additional contributions to the entropy. The heat map therefore presents the condensate value starting from a small but nonzero disorder value, which weakly lifts the degeneracy of the clean limit.}  
Overall, the entropy is low deep in
the normal and superradiant regions, while it attains a maximum in the condensed regime close to the mean-field instability line given by $\sqrt{\mu^2+\sigma^2}=\mu_c$ (marked as a red dashed curve). The entanglement map therefore gives an
independent finite-size signature of the same instability boundary found from
the photon number. 
Moreover, it shows that the strongest boson--fermion quantum correlations are localized near the finite-size transition region. On the flip side, far away from the instability there is little entanglement between the fermions and the bosonic mode and the mean-field treatment is expected to yield a good approximation. 

Taken together, the results of this section reveal two complementary regimes. Close to the phase boundary, the onset is governed by the weak-coupling criterion encoded in Eq.~\eqref{eq:a2_general_main}, for which $\mu$ and $\sigma$ enter primarily on an equivalent footing in the form $\mu^2+\sigma^2$. Farther inside the condensed phase, by contrast, the condensate is controlled by the spectral organization of the coupling matrix $g$---the arrangement of its eigenvalues $\{\lambda_m\}$, with density $\rho(\lambda)$, Eq.~\eqref{eq:Pure_disorder_rho}, whose occupied half fixes the polarization $k$ as per Eq.~\eqref{eq:k_strong_projector}, and the mean and disorder variance no longer play equivalent roles. Understanding the microscopic origin of this strong asymmetry requires examining how the fermionic state reorganizes in the condensed phase, which is the focus of the next section.

\section{Microscopic origin of disorder-enhanced superradiance}
\label{sec:mechanism}

\begin{figure}
    \centering
    \includegraphics[width=1.0\linewidth]{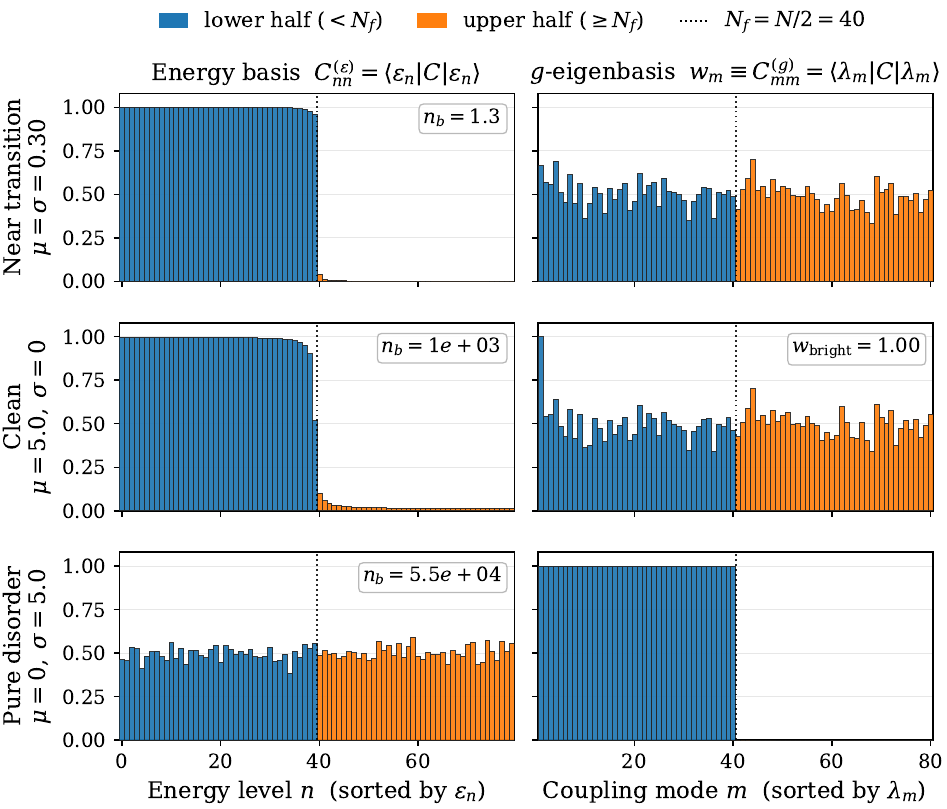}
    \caption{Microscopic structure of the condensed state across three representative regimes, shown for a single realization at $N=80$ ($N_f=40$). The left column shows the diagonal of the correlation matrix in the bare energy basis, $C^{(\varepsilon)}_{nn}=\langle \varepsilon_n|C| \varepsilon_n\rangle$, with levels ordered by increasing bare energy. The right column shows the diagonal of the same matrix in the eigenbasis of the coupling matrix, $w_m\equiv C^{(g)}_{mm}=\langle \lambda_m|C| \lambda_m\rangle$, with modes ordered by decreasing eigenvalue $\lambda_m$. 
    $N_f$ is denoted by a dotted line and blue (orange) bars mark states below (above) it. 
    From top to bottom the panels correspond to a point near the onset, $(\mu,\sigma)=(0.30,0.30)$, a point deep in the superradiant regime of the clean limit, $(\mu,\sigma)=(5.0,0)$, and one deep in the superradiant regime of pure disorder, $(\mu,\sigma)=(0,5.0)$.
    For the clean row, a small value $\sigma=10^{-3}$ is used to resolve the degenerate dark sector of $g$ for visualization of the right panel.
    Near onset, neither basis is strongly favoured. In the clean case, the occupation remains close to the free-Fermi step in the energy basis, while in the pure-disorder case it becomes close to a projector onto one half of the $g$-spectrum. Increasing disorder, therefore, shifts the organization of the condensed state from the bare-energy basis to the coupling eigenbasis.}
    \label{fig:mic-mech}
\end{figure}

To understand why disorder produces a parametrically stronger condensate, it is instructive to examine how the fermionic correlation matrix $C$ defined in Eq.~\eqref{eq:C_def} reorganizes across regimes. Figure~\ref{fig:mic-mech} shows $C$ in two complementary bases, the bare, zero-coupling basis $\{\ket{\epsilon_n}\}$ defined by the fermionic ladder, ordered by the single-particle energies $\epsilon_n$, and the eigenbasis $\{\ket{\lambda_m}\}$ of the coupling matrix $g$, ordered by its eigenvalues $\lambda_m$.

The left column displays the occupations $\langle \epsilon_n|C|\epsilon_n\rangle=\langle c_n^\dagger c_n\rangle$.
Close to the onset of the instability (top panel), the state is still dominated by the bare single-particle energies $\epsilon_n$ and the occupation profile remains close to a Fermi step function. Remarkably, the same structure persists deep in the superradiant phase of the clean model (middle panel), where only a narrow reconstruction appears around the Fermi surface. This shows that, in the clean model, the large degenerate dark sector of the coupling matrix leaves the many-body state essentially organized by the free-fermion spectrum even in the condensed phase.
The situation changes qualitatively in the pure-disorder regime (bottom panel). Here, the occupation profile becomes broadly smeared and nearly featureless across the harmonic ladder, indicating that the condensed state is no longer organized in the  $| \epsilon_n \rangle$ eigenbasis.

The right column reveals where this lost structure reappears. Near the onset of the instability (top row), the occupations remain weakly correlated with the $g$-spectrum, consistent with the fact that the many-body wavefunction is still determined by the normal state. Deep in the superradiant phase of the pure-disorder model (bottom right), a clear ordering emerges. The coupling term dominates the effective single-particle Hamiltonian and the $N_f$ eigenmodes with the largest (or smallest, depending on the sign of $\alpha$) eigenvalues $\lambda_m$ are almost fully occupied, showing a strong correlation between $C$ and $g$. This locking of $C$
to the spectrum of $g$ is the microscopic expression of superradiance as a coherent many-body effect. The fermionic state reorganizes so that extensively many grey eigenmodes participate cooperatively in the polarization $k=\mathrm{Tr}(gC)$
that sources the cavity field, rather than a single bright mode and extensively many dark modes of the clean model---which is precisely what makes the disordered condensate parametrically larger.

For the coupling matrix in the clean limit, instead, the dark sector of $g$ is $(N-1)$-fold degenerate, giving a large ambiguity in defining the basis states $| \lambda_m \rangle$. 
In Fig.~\ref{fig:mic-mech}, we therefore use a small disorder $\sigma=10^{-3}$ in the row representing the clean case to select a definite basis inside the dark sector\footnote{This perturbation has an insignificant effect on the scaling of the condensate and the occupation profile in the bare-energy basis.}. As the figure illustrates, there is a single strongly occupied bright mode, while the rest of the fermionic occupation is distributed across all other eigenstates $| \lambda_m \rangle$ of the coupling matrix.

The connection to superradiance becomes transparent by rewriting the fermionic polarization as
\begin{equation}
k=\mathrm{Tr}(gC)=\sum_m \lambda_m\, w_m,
\quad
w_m=\langle \lambda_m|C|\lambda_m\rangle.
\end{equation}
In the clean case, the coupling matrix contains one bright eigenvalue separated from an $(N-1)$-fold degenerate dark sector. In contrast, disorder broadens the dark sector into a continuum of distinct grey eigenvalues. In the condensed state, the occupied fermionic subspace aligns with one side of this spectrum, so that an extensive number of eigenmodes contribute coherently to $k$. The disordered condensate is therefore built from an $O(N)$ set of grey eigenmodes rather than from a single bright mode and extensively many dark modes, explaining its parametrically enhanced scaling.

We can make this mechanism more explicit in the strong-coupling regime by calculating the correlator between $g$ and $C$
\begin{equation}
  \mathrm{Corr}(g,C)
  :=\frac{\mathrm{Tr}(gC)}{\sqrt{\mathrm{Tr}(g^2)\,\mathrm{Tr}(C^2)}}\in[-1,1].
  \label{eq:corr_def}
\end{equation}
We now proceed by computing the traces of the matrices above. At $T=0$ and half filling, $C$ has rank $N/2$, hence $\mathrm{Tr}(C^2)=\mathrm{Tr}(C)=N/2$. 
In the clean limit, $g_{ij}=\mu(1-\delta_{ij})$, so
\begin{equation}
\mathrm{Tr}(g^2)=\mu^2 N(N-1),
\end{equation}
while in the pure-disorder limit, using $\rho(\lambda)$ as in Eq.~\eqref{eq:Pure_disorder_rho}, one obtains
\begin{equation}
\mathrm{Tr}(g^2)
\simeq N\int d\lambda \,\lambda^2\rho(\lambda)
=\sigma^2N(N-1).
\end{equation}
Together with the strong-coupling expressions for $k=\mathrm{Tr}(gC)$, this gives
\begin{equation}
  \mathrm{Corr}(g,C)\ \sim\
  \begin{cases}
    \pm\,\dfrac{1}{\sqrt{2(N-1)}}, &   \text{for} \, \sigma = 0,\\[8pt]
    \pm\,\dfrac{4\sqrt{2}}{3\pi}, & \text{for} \, \mu = 0,
  \end{cases}
  \label{eq:corr_results}
\end{equation}
where the sign reflects the choice of branch, i.e., whether the upper or lower half of the spectrum is filled.

Thus, in the clean case the fermionic correlations and the coupling matrix become asymptotically uncorrelated, while the pure-disorder case retains an  $O(1)$ alignment between the coupling and fermionic correlation matrices. In the language of Fig.~\ref{fig:mic-mech}, this means that the condensed state in the clean model remains organized primarily by the bare-energy basis, whereas the pure-disorder state is reorganized into an almost ideal projector onto the leading part of the $g$-spectrum. Microscopically, the disorder-dominated condensate is therefore the result of the alignment between the occupied subspace and an extensive set of coupling-matrix  eigenmodes.

Using Eq.~\eqref{eq:rho_general}, we can also calculate the correlation at generic values of $\mu$ and $\sigma$, which at large-$N$ is given by the expression
\begin{equation}
    \mathrm{Corr}(g,C)\ \sim \pm \frac{1}{\sqrt{2 N}} \frac{\mu}{\sqrt{\mu^2+\sigma^2}} \pm \frac{4 \sqrt{2}}{3 \pi} \frac{\sigma}{\sqrt{ \mu^2+\sigma^2}} \, .
\end{equation}
At any non-vanishing $\sigma$, the strong-coupling correlations are dominated by the disorder at large $N$ and remain finite in the thermodynamic limit. 

\section{Conclusion}
\label{sec:conclusion}

We have studied an idealized model for cold fermionic atoms in an optical cavity, described by a fermion--boson model in which $N$ internal levels arranged as a harmonic ladder couple to a single cavity mode through a real symmetric matrix $g_{ij}$. The mean-field saddle reduces to a single self-consistency condition $k = \mathrm{Tr}(gC)$, which encodes how the correlations in the occupied fermionic subspace align with the coupling matrix. Based on the Landau instability criterion of the mean-field free energy, and corroborated by exact diagonalization, we showed that the onset of superradiance is governed by the criterion $\mu^2+\sigma^2=\mu_c^2$, in which disorder variance plays the same role as the mean coupling strength. 

This equivalence is, however, a threshold property only. In the condensed phase, the bosonic field amplitude in the clean model scales as $|\alpha_\ast|\propto\sqrt{N}$, while the addition of disorder leads to the parametrically stronger scaling $|\alpha_\ast|\propto N$. 
As we have argued, the reason lies in the fact that the coupling matrix in the clean model supports a single bright eigenmode and an extensive number of dark modes, while in the disorder-dominated regime an $O(N)$ set of grey eigenmodes of the coupling matrix contributes coherently to the fermionic polarization. 
We quantified this mechanism through the normalized correlator between the coupling matrix and the fermionic correlation functions, $\mathrm{Corr}(g,C)$. It decays as $1/\sqrt{N}$ in the clean model while it is $O(1)$ in the disordered phase, showing a strong alignment between the couplings and the fermionic modes. 

Moreover, we have been able to analytically justify the drastically distinct scaling behavior resorting to random-matrix theory, which faithfully describes the spectrum of the coupling matrix $g$ in the strong-coupling regime and at large $N$. The semicircle law, supplemented by the clean bright outlier, allowed us to evaluate the occupied spectral weight of grey modes entering the fermionic polarization $k=\mathrm{Tr}(gC)$ and thereby derive the analytic interpolation between the Dicke-like clean scaling and the disorder-dominated $|\alpha_\ast|\propto N$ regime. 

We have validated these predictions across the $(\mu,\sigma)$ plane by numerically solving the full mean-field equations and by exact diagonalization at small system sizes. 
In addition, the numerically exact boson--fermion entanglement entropy shows that quantum correlations are concentrated near the finite-size transition region and remain weak deep in the normal and condensed regimes, rendering the mean-field description quantitatively reliable anywhere in the parameter space except close to the superradiance onset.

Our derivation has assumed that the coupling matrix $g_{ij}$ is drawn from the GOE, yet this is not a fundamental constraint: the essential ingredient is the existence of an 
extensive number of grey eigenmodes that coherently polarize the fermionic manifold. 
The semicircle spectral density, and hence our analytic  scaling argument, extends without modification to any real symmetric random matrix with 
i.i.d.\ entries of zero mean and finite variance, irrespective of the microscopic 
distribution~\cite{Mehta2004}. 
Two further physically motivated departures from the GOE preserve the semicircular bulk: 
sparsifying the coupling matrix, as can arise from geometrically constrained 
atom--cavity interactions, yields a semicircle law whenever the average connectivity grows 
with $N$~\cite{Rodgers1988,K_hn_2008}; and imposing an algebraically decaying distance 
dependence $g_{ij}\sim|i-j|^{-\beta}$ places the matrix in the power-law random banded 
matrix class~\cite{mirlin1996}, where the bulk spectral density remains 
approximately semicircular in the delocalized phase. 
We therefore expect the disorder-enhanced scaling $|\alpha_\ast|\propto N$ to be a robust 
feature of the condensed phase, not an artifact of Gaussian disorder.

Several natural extensions emerge from this work. First, realistic cavity platforms operate in the driven-dissipative regime, where cavity photon loss and atomic decay enter through Lindblad or non-Markovian bath couplings~\cite{EmaryBrandes2003,Dimer2007}. Dissipation generically shifts the superradiant critical point to larger coupling values~\cite{Nagy:2011bok}, and it is an open question whether the disorder-enhanced field $|\alpha_\ast|\propto N$ survives under photon loss or is replaced by a different nonequilibrium steady state. Second, extending the single-mode setup to a multimode cavity~\cite{Gopalakrishnan2009,Gopalakrishnan2011,Vaidya2018} would allow the coupling matrix to become a full random tensor, opening a route toward a glassy superradiant phase. In multimode cavity QED, frustrated and disordered interactions have already been studied in the bosonic case \cite{Strack:2011ggg} and experimentally engineered~\cite{Sauerwein2023}, and replica symmetry breaking has been directly observed in driven-dissipative spin glasses~\cite{Marsh:2023odt,Kroeze:2023thu}. It remains an open problem whether a fermionic internal manifold can support a coexistence of superradiant and spin-glass order---an analogue of the charge-glass proposals for itinerant fermions~\cite{Glassy_physics}. Third, the boson--fermion coupling structure studied here coincides with that of the Yukawa-SYK model in the single-mode limit~\cite{Esterlis_2019,YSYK_SelfTunned_QCrit,PascualSolis2026YSYK}. Retaining quantum fluctuations beyond mean field would determine whether the critical cavity photons at the superradiant onset induce non-Fermi-liquid behavior~\cite{Mandal:2024uxx}, as has been argued for fermionic lattice models near the self-organization transition. We hope this work motivates the deeper exploration of disorder as a resource for collective light--matter phenomena beyond the Dicke paradigm.

\section*{Author contributions}

D.P.S. and A.L. performed the analytical calculations. D.P.S. performed the numerical simulations. All authors contributed to the conception of the project, the interpretation of the results, and the writing of the manuscript. During preparation, AI-assisted tools were used as auxiliary support for editing and coding. All authors discussed the results and approved the final manuscript.

\section*{Acknowledgments}
This project has been funded by the Caritro Foundation and by the Swiss State Secretariat for Education, Research and Innovation (SERI) under contract number UeMO19-5.1. This work was supported by the Provincia Autonoma di Trento, and Q@TN, the joint lab between the University of Trento, FBK—Fondazione Bruno Kessler, INFN—National Institute for Nuclear Physics, and CNR—National Research Council, Italy.

\bibliographystyle{apsrev4-2}
\bibliography{references}

\clearpage
\newpage
\onecolumngrid

\appendix

\section{Curvature at $\alpha=0$ and the onset criterion on the ladder}
\label{app:landau}

This appendix derives the curvature coefficient $a_2=\tfrac{1}{2}F''(0)$,
introduced in Eq.~\eqref{eq:a2}, which determines the local stability of the normal solution $\alpha=0$
for the harmonic ladder at $T=0$ and half filling $N_f = N/2$.
As in the main text, we assume $g_{ii}=0$.

At fixed order parameter $\alpha$, the mean-field single-particle Hamiltonian is
\begin{equation}
  h(\alpha) = \epsilon + V(\alpha),
  \qquad
  \epsilon_{ij} = \epsilon_i\,\delta_{ij},
  \qquad
  \epsilon_i = \omega_{\mathrm{f}}\!\left(i - \frac{N+1}{2}\right),
\end{equation}
where 
\begin{equation}
  V_{ij}(\alpha) = 2\sqrt{\frac{2}{N}}\,\alpha\,g_{ij}.
  \label{eq:V_app}
\end{equation}
is effectively a perturbation for small $\alpha$, while the unperturbed single-particle energies $\epsilon_i$ are the
levels of the harmonic ladder.

The mean-field free energy at $T=0$ is
\begin{equation}
  F(\alpha) = \omega_0\,\alpha^2 + \sum_{i=1}^{N_f} E_i(\alpha),
  \label{eq:F_T0_app}
\end{equation}
where $E_i(\alpha)$ are the eigenvalues of $h(\alpha)$ sorted in increasing order.
We expand each eigenvalue $E_i(\alpha)$ in powers of $\alpha$ around $\alpha = 0$ using perturbation theory,
\begin{equation}
  E_i(\alpha) = \epsilon_i + \langle i|V|i\rangle +  \sum_{\substack{j=1\\j\neq i}}^{N}
      \frac{|\langle j|V|i\rangle|^2}{\epsilon_i - \epsilon_j} + O(\alpha^3).
\end{equation}
Since $V_{ii}(\alpha) \propto g_{ii} = 0$, the first-order contribution to $F$ vanishes identically and we are left with the second-order term only:
\begin{equation}
  F(\alpha) = F(0) + \alpha^2
  \left[
    \omega_0 + \frac{8}{N}
    \sum_{i=1}^{N/2}\,
    \sum_{\substack{j=1\\j\neq i}}^{N}
    \frac{g_{ij}^2}{\epsilon_i - \epsilon_j}
  \right]
  + O(\alpha^4),
  \label{eq:F_expand_app}
\end{equation}
where we substituted the explicit expression for $V(\alpha)$.

The second-order derivative of the free energy therefore reads
\begin{equation}
  a_2(g)=\omega_0+\frac{8}{N}\sum_{i=1}^{N/2}\sum_{j\neq i}\frac{g_{ij}^2}{\epsilon_i-\epsilon_j},
  \label{eq:a2_exactsum_app}
\end{equation}
or, writing explicitly the harmonic ladder energy levels $\epsilon_i-\epsilon_j=\omega_{\mathrm f}(i-j)$
\begin{equation}
  a_2(g)=\omega_0+\frac{8}{N\omega_{\mathrm f}}
  \sum_{i=1}^{N/2}\sum_{j\neq i}\frac{g_{ij}^2}{i-j}.
  \label{eq:a2_ladder_app}
\end{equation}
In order to obtain a single onset estimate that covers clean, disordered, and mixed couplings, we use the fact that each off-diagonal entry $g_{ij}$ has the same mean $\mu$ and variance $\sigma^2$. In the
double sum in \eqref{eq:a2_ladder_app}, $g_{ij}^2$ can be replaced by its typical size,
\begin{equation}
  g_{ij}^2\ \longrightarrow\ \mu^2+\sigma^2,
  \qquad (i\neq j).
  \label{eq:replace_second_moment_app}
\end{equation}
For the clean all-to-all model this replacement is exact since $g_{ij}^2=\mu^2$ off-diagonal, while
for the Gaussian ensemble it is an application of the central limit theorem for independent entries $|g_{ij}|^2$ at fixed distance $i-j$.
Using this property, the sum of the energy levels can be explicitly performed. By defining
\begin{equation}
  S_N:=\sum_{i=1}^{N/2}\sum_{j\neq i}\frac{1}{i-j}
  \label{eq:SN_def_app}
\end{equation}
and introducing the harmonic numbers $H_n=\sum_{m=1}^n 1/m$, which satisfy 
\begin{equation}
  \sum_{j\neq i}\frac{1}{i-j}=H_{i-1}-H_{N-i},
  \label{eq:inner_harmonic_app}
\end{equation}
we can write $S_N=\sum_{i=1}^{N/2}(H_{i-1}-H_{N-i})$. In the large $N$ limit, and for $N$ even
\begin{equation}
  S_N=-\,N\log 2+o(N),
  \qquad N\to\infty.
  \label{eq:SN_asympt_app}
\end{equation}

Combining Eq.~\eqref{eq:replace_second_moment_app} and Eq.~
\eqref{eq:SN_asympt_app}, Eq.~\eqref{eq:a2_ladder_app} becomes
\begin{equation}
  a_2 = \omega_0-\frac{8\log 2}{\omega_{\mathrm f}}(\mu^2+\sigma^2),
  \qquad (N\to\infty,\ N\ \mathrm{even}).
  \label{eq:a2_mu_sigma_app}
\end{equation}
The onset is given by $a_2=0$, which yields
\begin{equation}
  \mu^2+\sigma^2= \frac{\omega_0\omega_{\mathrm f}}{8\log 2}.
  \label{eq:critical_circle_app}
\end{equation}

Equation \eqref{eq:critical_circle_app} contains the clean and disordered limits as special cases. Setting $\sigma=0$ gives the clean threshold $\mu_c=\sqrt{\omega_0\omega_{\mathrm f}/(8\log 2)}$ while setting $\mu=0$ gives the pure-disorder threshold $\sigma_c=\sqrt{\omega_0\omega_{\mathrm f}/(8\log 2)} = \mu_c$. These values are used in Sec.~\ref{sec:clean_baseline} and Sec.~\ref{sec:pure_disorder} as the reference onset scales for the clean and pure-disorder cuts. They are the two axis intercepts of the circular boundary in Eq.~\eqref{eq:critical_circle}, which is shown as the red dashed arc in Fig.~\ref{fig:contour}. The same thresholds are tested in Fig.~\ref{fig:mf_summary}(b,d), where the zero crossings of the Landau coefficient converge to the analytical prediction, and in Fig.~\ref{fig:ed_mf}, where exact diagonalization reproduces the same onset scale at small system sizes.

\section{Dicke limit and comparison with the full fermion--boson model}
\label{app:dicke_limit}

The connection between the clean model discussed in section \ref{sec:clean_baseline} and the Dicke model extends beyond a simple phenomenological analogy. The fermionic Hamiltonian
\eqref{eq:H} contains a Dicke model as a particular limit, and this
embedding helps separate two ingredients of the problem. One is the collective bright-mode mechanism that is already familiar from Dicke physics \cite{Dicke1954,HEPP1973360,WangHioe1973,EmaryBrandes2003,Kirton2019}. The other is the additional matrix reorganization that becomes possible in the full fermion--boson model, especially in the presence of disorder. The purpose of this appendix is to make that distinction explicit.

The starting point is the harmonic ladder \eqref{eq:ladder}. For even $N$, the
spectrum is symmetric around zero, so each level $i=1,\dots,N/2$ has a natural
partner $\bar i=N+1-i$ with
\begin{equation}
  \epsilon_{\bar i}=-\epsilon_i .
\end{equation}
The finite single-particle manifold can therefore be viewed as a set of $N/2$ particle-hole doublets. If one restricts the Hilbert space to the sector with exactly one fermion in each doublet,
\begin{equation}
  c_i^\dagger c_i + c_{\bar i}^\dagger c_{\bar i}=1 ,
  \label{eq:Dicke_doublet}
\end{equation}
then each pair behaves as an effective two-level system. Introducing
pseudospin operators
\begin{equation}
  S_i^z=\frac{1}{2}\left(c_{\bar i}^\dagger c_{\bar i}-c_i^\dagger c_i\right),
\end{equation}
\begin{equation}
  S_i^x=\frac{1}{2}\left(c_{\bar i}^\dagger c_i+c_i^\dagger c_{\bar i}\right),
\end{equation}
the fermionic energy of the pair becomes
\begin{equation}
  \epsilon_i c_i^\dagger c_i+\epsilon_{\bar i} c_{\bar i}^\dagger c_{\bar i}
  = \Omega_i S_i^z ,
\end{equation}
with level splitting
\begin{equation}
  \Omega_i=\epsilon_{\bar i}-\epsilon_i
  =\omega_{\mathrm f}(N+1-2i).
\end{equation}
At this stage the fermionic model has been rewritten as a set of two-level
systems with generally nonuniform splittings $\Omega_i$.

To obtain a Dicke Hamiltonian one must further constrain the coupling matrix.
Assume that $g_{ij}$ connects only the two states within each doublet,
\begin{equation}
  g_{ij}=0 \qquad \text{unless } j=\bar i ,
\end{equation}
and write the nonzero matrix elements as
\begin{equation}
  g_{i\bar i}=g_{\bar i i}=\lambda_i .
\end{equation}
Then, the interaction term in Eq.~\eqref{eq:H} reduces to
\begin{equation}
  \sqrt{\frac{2}{N}}(a+a^\dagger)
  \sum_{i=1}^{N/2}\lambda_i
  \left(c_{\bar i}^\dagger c_i+c_i^\dagger c_{\bar i}\right)
  =
  \frac{2}{\sqrt{N/2}}(a+a^\dagger)\sum_{i=1}^{N/2}\lambda_i S_i^x .
\end{equation}
The constrained Hamiltonian is therefore
\begin{equation}
  H_{\mathrm{pair}}
  =
  \omega_0 a^\dagger a
  +\sum_{i=1}^{N/2}\Omega_i S_i^z
  +\frac{2}{\sqrt{N/2}}(a+a^\dagger)\sum_{i=1}^{N/2}\lambda_i S_i^x .
  \label{eq:H_pair_dicke}
\end{equation}
Equation \eqref{eq:H_pair_dicke} is an inhomogeneous Dicke model written in the
same normalization conventions as the main text. The standard Dicke limit is
recovered when all doublets have the same splitting and the same coupling,
\begin{equation}
  \Omega_i=\Omega,
  \qquad
  \lambda_i=\lambda ,
\end{equation}
in which case
\begin{equation}
  H_{\mathrm D}
  =
  \omega_0 a^\dagger a
  +\Omega S^z
  +\frac{2\lambda}{\sqrt{N/2}}(a+a^\dagger)S^x ,
\end{equation}
with $S^\alpha=\sum_{i=1}^{N/2}S_i^\alpha$.

This mapping is useful because it pinpoints the precise Dicke-like structure embedded in the fermionic problem. The clean model discussed in the main text has a coupling matrix that acts over the full finite single-particle manifold rather than in the form of the pairwise interactions of Eq.~\eqref{eq:Dicke_doublet}, and is therefore not exactly equivalent to the Dicke model. Nevertheless, its spectrum still features a single bright collective mode separated from a dark sector. This structural property explains why the clean model exhibits the same mean-field superradiant phenomenology as the Dicke model. The continuous onset, the two symmetry-related branches, and the scaling $|\alpha_\ast|\propto\sqrt{N}$ all arise from this collective bright-mode mechanism.

\begin{figure*}[t!]
    \centering
    \includegraphics[width=1.0\linewidth]{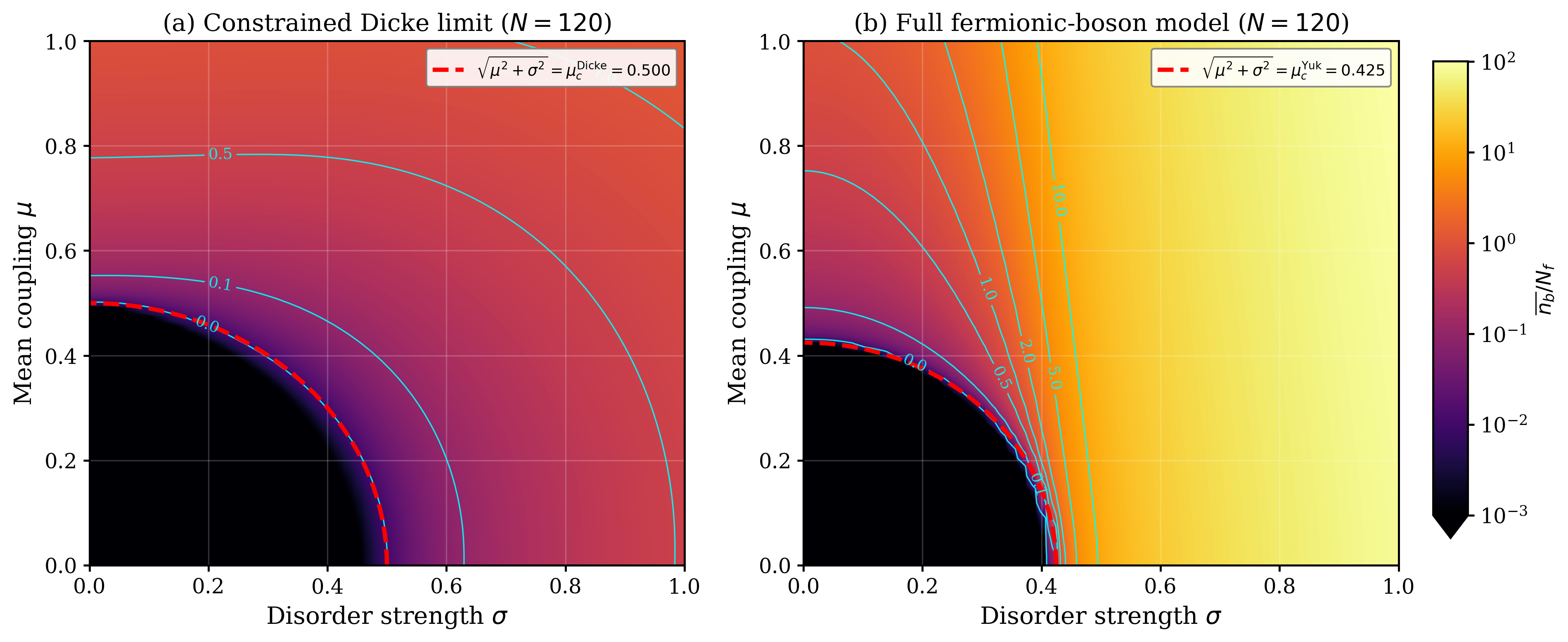} 
    \caption{Comparison between the constrained Dicke limit and the full fermion--boson model at $N=120$. (a) Normalized boson occupation in the constrained Dicke limit, obtained from the constrained pairing construction described in App.~\ref{app:dicke_limit}. (b) Corresponding disorder averaged result from the full fermion--boson model. In both panels, the red dashed line marks the onset estimate $\sqrt{\mu^2+\sigma^2}=\mu_c$, and the cyan curves are iso-$n_b$ contours. The clean sectors are organized in a similar Dicke-like manner, but the effect of disorder is qualitatively different. In the constrained Dicke limit, disorder only reshuffles effective two-level couplings, whereas in the full model it reorganizes the internal manifold and produces a much stronger enhancement of the condensate.}
    \label{fig:dicke_vs_yukawa}
\end{figure*}

The heatmaps in Fig.~\ref{fig:dicke_vs_yukawa} make this comparison explicit.
Panel~(a) shows the bosonic-mode population in the constrained Dicke limit, where disorder acts only through the
effective two-level couplings $\lambda_i$. Panel~(b) shows the analogous values for the full fermion--boson model,
where disorder enters the full matrix ensemble \eqref{eq:g_disorder}. In both
cases, the onset is well organized by the same estimate
$\sqrt{\mu^2+\sigma^2}=\mu_c$, which reflects the weak-coupling argument
discussed in the main text. The difference appears deeper in the condensed
region. In the constrained Dicke limit, the contour pattern remains closer to the circular geometry of the onset criterion, because disorder only redistributes the strengths of otherwise independent collective channels, as in disordered Dicke models \cite{Das2024}. In the full model, by contrast, disorder acts on the full matrix spectrum, whose large-$N$ form is governed by the semicircle law used in Appendix~\ref{app:disorder_strong}. This is why the contour lines bend much more strongly and tend to become parallel to the $\mu$ axis.

The comparison therefore clarifies the scope of the Dicke analogy. It captures
the clean bright-mode mechanism, while the disorder enhancement relies on the spectral structure of the full coupling matrix.

\section{Details on the strong coupling limit}

In this appendix, we analyze the regime in which the cavity-induced term dominates
over the harmonic ladder in the mean-field single-particle Hamiltonian. Starting
from the mean-field Eq.~\eqref{eq:HMF_main}, 
we assume that, sufficiently far inside the condensed phase, the contribution
proportional to $g$ sets the leading structure of the spectrum, while the ladder
term $\epsilon_i$ only provides a subleading correction. In this strong-coupling
limit, one may therefore approximate
\begin{equation}
  h(\alpha)\approx 2\sqrt{\frac{2}{N}}\,\alpha\,g.
\end{equation}
The occupied fermionic subspace is then selected primarily by the eigenstructure
of the coupling matrix $g$, which makes it possible to derive explicit
large-$N$ expressions for the fermionic polarization $k=\mathrm{Tr}(gC)$ and for the
associated condensate observables. We first discuss the clean limit, then the
pure-disorder case, and finally the general scenario with both finite mean and
variance.

\subsection{Clean model}
\label{app:clean_fourier}

To derive the results presented in Sec.~\ref{sec:clean_baseline}, we start from the clean model interaction
\begin{equation}
g_{ij}=\mu(1-\delta_{ij}),\qquad i,j=1,\dots,N.
\label{eq:app_g_clean}
\end{equation}
Notice that $g$ is a circulant matrix. Each row is obtained from the previous one by a cyclic shift or, equivalently, $g_{ij}$ depends
only on $i-j$ modulo $N$.

Circulant matrices are diagonalized by a discrete Fourier transform. For $q=0,1,\dots,N-1$, we define
\begin{equation}
|q\rangle=\frac{1}{\sqrt N}\sum_{j=1}^N
\exp\!\left(i\frac{2\pi q}{N}(j-1)\right)\,|j\rangle,
\label{eq:app_fourier_vecs}
\end{equation}
where $q$ labels the Fourier mode. The eigenvalues of $g$ can be read from the Fourier transform of the first row $(0,\mu,\mu,\dots,\mu)$
\begin{equation}
\lambda_q=\mu\sum_{m=1}^{N-1}\exp\!\left(-i\frac{2\pi q}{N}m\right).
\label{eq:app_lambdaq}
\end{equation}
For $q=0$, the sum equals $N-1$, so $\lambda_0=\mu(N-1)$. For $q\neq0$, we use
$\sum_{m=0}^{N-1}e^{-i2\pi q m/N}=0$, hence $\sum_{m=1}^{N-1}e^{-i2\pi q m/N}=-1$, and therefore
\begin{equation}
\lambda_0=\mu(N-1),\qquad \lambda_q=-\mu \ \ (q=1,\dots,N-1).
\label{eq:app_clean_spectrum}
\end{equation}
The spectrum contains one extensive eigenvalue and $N-1$ degenerate $O(1)$ eigenvalues. As we will see presently, we can interpret the former as a bright mode, while the latter are dark.  

We can express the fermionic operators in the Fourier basis as $d_q=\sum_j\langle j|q\rangle\,c_j$, and denote
$n_q=\langle d_q^\dagger d_q\rangle$. Then
\begin{equation}
k=\mathrm{Tr}(gC)=\sum_{q=0}^{N-1}\lambda_q\,n_q,
\qquad
\sum_{q=0}^{N-1}n_q=N_f. 
\label{eq:app_k_fourier}
\end{equation}
The fermionic polarization determines the strength of the cavity field via the self-consistency in the main text, Eq.~\eqref{eq:alpha_k},
\begin{equation}
\alpha=-\sqrt{\frac{2}{N}}\frac{k}{\omega_0}. 
\end{equation}
The eigenmode $\lambda_0$ yields a macroscopic cavity field that scales $\sim \sqrt{N}$, and is therefore a collective bright mode. 
In contrast, the eigenmodes $\lambda_q,\ q>0$ contribute each a cavity field with amplitude $\sim 1/\sqrt{N}$, and are therefore (almost) dark. They can nevertheless generate a significant contribution, since at half filling there will be an extensive number of these dark modes occupied.  

To determine $k$, we just need to understand whether the bright mode is among the occupied states. In the strong-coupling limit, the single-particle energies are proportional to $\alpha\lambda_q$, where $\lambda_q$ are the eigenvalues of $g$. Thus, for $\mu>0$, the sign of $\alpha$ determines whether the bright eigenvalue $\lambda_0=\mu(N-1)$ lies at the top or at the bottom of the spectrum. If $\alpha > 0$, the bright mode is empty, and $N/2$ degenerate dark modes are filled, yielding
\begin{equation}
k=(-\mu)\frac{N}{2}=-\frac{\mu N}{2}.
\label{eq:app_k_minus}
\end{equation}
For $\alpha < 0$, the bright mode has the lowest single-particle energy, while the remaining $N/2-1$ occupied modes are dark and
\begin{equation}
k=\mu(N-1)-\mu\left(\frac{N}{2}-1\right)=\frac{\mu N}{2}.
\label{eq:app_k_plus}
\end{equation}
Thus, $k=\pm \mu N/2$ to leading order in $N$. The two signs correspond to the two $\mathbb{Z}_2$-related branches of the mean-field solution.

Using the self-consistency Eq.~\eqref{eq:alpha_k},
we obtain
\begin{equation}
\alpha=\mp\,\frac{\mu}{\omega_0}\sqrt{\frac{N}{2}},
\qquad
n_b=\alpha^2=\frac{\mu^2}{\omega_0^2}\frac{N}{2}.
\label{eq:app_alpha_nb_scaling}
\end{equation}
This gives the $\sqrt{N}$ growth of $|\alpha|$ and the linear growth of $n_b$ in the clean
all-to-all model. The value of $\mu_c$ is set by the competition with the ladder spectrum
and is treated in the main text.

\subsection{Pure-disorder regime}
\label{app:disorder_strong}

This appendix derives the disorder-dominated scaling quoted in Sec.~\ref{sec:pure_disorder} of the main text. Let $\lambda_1\geq\lambda_2\geq\cdots\geq\lambda_N$ be the eigenvalues
of $g$ with eigenvectors $|\lambda_m\rangle$. For large $|\alpha|$, the eigenvalues of $h(\alpha)$ are
approximately $2\sqrt{2/N}\,\alpha\,\lambda_m$. At $T=0$ and half filling, the ground state
fills the $N/2$ lowest single-particle energies. Choosing the branch with $\alpha<0$, the lowest single-particle energies correspond to the largest eigenvalues of $g$. Choosing $\alpha>0$ instead fills the opposite side of the spectrum and gives the same value of $|\alpha|$ and $n_b$ in the large-$N$ semicircle approximation, or after disorder averaging. The correlation matrix therefore approaches the projector
\begin{equation}
  C\ \approx\ \sum_{m=1}^{N/2}|\lambda_m\rangle\langle \lambda_m|.
\end{equation}
Consequently,
\begin{equation}
  k=\mathrm{Tr}(gC)\ \approx\ \sum_{m=1}^{N/2}\lambda_m.
  \label{eq:app_k_top_half}
\end{equation}
The condensate then follows from the self-consistency relation derived in the
main text, Eq.~\eqref{eq:alpha_k}, so that
\begin{equation}
  \alpha_\ast=-\sqrt{\frac{2}{N}}\frac{k(\alpha_\ast)}{\omega_0},
  \qquad
  n_b=\alpha_\ast^2=\frac{2}{N}\frac{k(\alpha_\ast)^2}{\omega_0^2}.
  \label{eq:app_nb_from_k}
\end{equation}

In the pure disorder case $\mu=0$, we can write $g=\sigma\xi$, with
\begin{equation}
  \xi_{ij}=\xi_{ji},\qquad \xi_{ii}=0,\qquad \mathbb E[\xi_{ij}]=0,\qquad \mathbb E[\xi_{ij}^2]=1
  \quad (i\neq j).
  \label{eq:app_xi_stats}
\end{equation}
In the large $N$ limit, the spectral density is described by the Wigner semicircle centered at $0$, as expected for a GOE\footnote{Notice that $\xi$ is not strictly a GOE matrix, since it has vanishing diagonal components. However, this just produces a correction that vanishes in the large-$N$ limit.} matrix \cite{Mehta2004,AGZ2010,Forrester2010}, 
\begin{equation}
  \rho_R(\lambda)=\frac{2}{\pi R^2}\sqrt{R^2-\lambda^2},
  \qquad |\lambda|\le R,
  \label{eq:app_semicircle}
\end{equation}
normalized as $\int_{-R}^R\rho_R(\lambda)\,d\lambda=1$. The radius $R$ follows from second-moment matching.
The second moment for a matrix $M$ distributed as in Eq.~\eqref{eq:app_semicircle} is
\begin{equation}
  \frac{1}{N}\,\mathrm{Tr}(M^2)\ \approx\ \int_{-R}^R \lambda^2 \rho_R(\lambda)\,d\lambda
  = \frac{R^2}{4}.
  \label{eq:app_moment_match_generic}
\end{equation}
Applying Eq.~\eqref{eq:app_moment_match_generic} to $M=\sigma\xi$, using
\begin{equation}
  \mathbb E\,\mathrm{Tr}(\xi^2)=\sum_{i\neq j}\mathbb E[\xi_{ij}^2]=N(N-1),
\end{equation}
gives
\begin{equation}
  R=2\sigma\sqrt{\frac{1}{N}\,\mathbb E\,\mathrm{Tr}(\xi^2)}
  =2\sigma\sqrt{N-1}.
  \label{eq:app_R_full}
\end{equation}
The eigenvalues thus scale as $\sqrt{N-1}$, in between the scaling of the bright mode ($\sim N$) and the dark modes ($O(1)$). Following Eq.~\eqref{eq:app_k_top_half} and the self-consistency Eq.~\eqref{eq:alpha_k}, fermionic modes with such a scaling contribute a cavity field amplitude of $O(1)$ and can therefore be interpreted as grey. The coherent addition of many such grey modes is what generates the disorder-enhanced cavity field, as we will show now.  

By symmetry, half of the eigenvalues are positive and half are negative, and therefore we can obtain Eq.~\eqref{eq:k_strong_projector} by integrating over the positive half of the semicircle.
For the branch chosen above, this gives
\begin{equation}
k \approx N\int_0^R d\lambda\,\lambda\rho_R(\lambda)
=\frac{4\sigma}{3\pi}\,N\sqrt{N-1},
\end{equation}
up to the sign fixed by the choice of $\alpha$. Inserting this result into
Eq.~\eqref{eq:app_nb_from_k} yields
\begin{equation}
  |\alpha|\approx \frac{4\sqrt{2}}{3\pi}\frac{\sigma}{\omega_0}\sqrt{N(N-1)},
  \qquad
  n_b\approx \frac{32}{9\pi^2}\frac{\sigma^2}{\omega_0^2}\,N(N-1).
  \label{eq:app_nb_pure_disorder}
\end{equation}
These expressions are the strong-disorder estimates quoted in
Eq.~\eqref{eq:pure_disorder_scaling_main} of the main text. They are used as the dashed asymptotic prediction for the pure-disorder cut in Fig.~\ref{fig:mf_summary}(c), where the numerical mean-field data approach the predicted $|\alpha_\ast|\propto N$ scaling away from the onset. The same
scaling also provides the pure-disorder limit of the general strong-coupling expression discussed in Sec.~\ref{sec:general_case}. 

\subsection{General case}
\label{app:generic_case}

We now consider the general case discussed in Sec.~\ref{sec:general_case}, where we have both finite mean and disorder
\begin{equation}
  g_{ij}=\mu(1-\delta_{ij})+\sigma\,\xi_{ij},
  \label{eq:app_g_mu_sigma_xi}
\end{equation}
or, in matrix terms
\begin{equation}
    g = \mu (N \ket{u}\bra{u} - I)+ \sigma \xi \, .
\end{equation}
Here, $I$ is the identity matrix and we have introduced the vector
\begin{equation}
  \ket{u}=\frac{1}{\sqrt N}(1,\dots,1)^T,
\end{equation}
which is the bright state of the clean model discussed in Sec.~\ref{app:clean_fourier}, with eigenvalue
$\mu(N-1)$. The mixed model can be viewed as a deterministic rank-one deformation of a Wigner matrix, together with a uniform shift in the sector orthogonal to the bright mode
\cite{CapitaineDonatiMartinFeral2009,KnowlesYin2013}.
Notice that, in the limit $N \to \infty$, $\ket{u}$ becomes an eigenvector of $g$ again with eigenvalue $\mu(N-1)$, 
\begin{equation}
    g \ket{u} = \mu(N-1) \ket{u} + \sigma \xi  \ket{u} \to \mu(N-1) \ket{u} \, ,
\end{equation}
as a consequence of the central limit theorem, since each component of $\xi  \ket{u}$ is the sum over $N$ independent Gaussian random variables with zero mean.

Now that we have determined the approximate bright state in the large $N$ limit, we can project $g$ on the subspace orthogonal to the bright state, namely the clean dark sector, using
\begin{equation}
  P=I- \ket{u}\bra{u} \, .
\end{equation}
We have
\begin{equation}
    PgP = \sigma P \xi P - \mu P.
    \label{eq:projected_g}
\end{equation}
Since $\xi  \ket{u} \sim 0$, we have $P \xi P \sim \xi$, which means that in the thermodynamic limit the disorder part of the interaction is invariant when projected on the orthogonal clean-dark subspace, and therefore it is still a GOE matrix with zero diagonal components. This suggests that the eigenvalue distribution is still given by the Wigner semicircle on this subspace. The second term of Eq.~\eqref{eq:projected_g} is nothing but the identity matrix when restricted to the subspace, which therefore shifts the distribution by a constant $-\mu$. As a result, we have the following eigenvalue distribution, which becomes exact in the large-$N$ limit:
\begin{equation}
    \rho(\lambda) = \frac{N-1}{N} \frac{2}{\pi R^2} \sqrt{R^2-(\lambda+\mu)^2} + \frac{1}{N} \delta(\lambda - \mu(N-1)) \, ,
\end{equation}
where $R$ can be determined again using Eq.~\eqref{eq:app_moment_match_generic} with
$M=\sigma P\xi P$ and dimension $N-1$:
\begin{equation}
  R
  =2\sigma\sqrt{\frac{1}{N-1}\,\mathbb E\,\mathrm{Tr}\!\big((P\xi P)^2\big)}.
  \label{eq:app_R_dark_from_moment}
\end{equation}
It remains to evaluate the trace expectation. Using $P^2=P$ and cyclicity of the trace, we can write $
  \mathrm{Tr}\!\big((P\xi P)^2\big)=\mathrm{Tr}(P\xi P\xi)$, which can be directly evaluated expanding $P$:
\begin{equation}
  \mathrm{Tr}(P\xi P\xi)
  =\mathrm{Tr}(\xi^2)-2\,\bra{u}\xi^2\ket{u}+(\bra{u}\xi \ket{u})^2.
  \label{eq:app_trace_identity_proj}
\end{equation}
The three expectations are elementary for the zero-diagonal ensemble as given in Eq.~\eqref{eq:app_xi_stats}:
\begin{align}
  \mathbb E\,\mathrm{Tr}(\xi^2)
  &= \sum_{i\neq j}\mathbb E[\xi_{ij}^2]
   = N(N-1), \label{eq:app_trace_xi2_exp} \\
  \mathbb E(\bra{u}\xi^2\ket{u})
  &= \mathbb E\|\xi u\|^2
   = \frac{1}{N}\sum_{i=1}^N\sum_{j\neq i}\mathbb E[\xi_{ij}^2]
   = N-1, \label{eq:app_ut_xi2_u_exp} \\
  \mathbb E(\bra{u}\xi \ket{u})^2
  &= \frac{4}{N^2}\sum_{i<j}\mathbb E[\xi_{ij}^2]
   = \frac{2(N-1)}{N}. \label{eq:app_ut_xi_u_sq_exp}
\end{align}
Substituting Eqs.~\eqref{eq:app_trace_xi2_exp}--\eqref{eq:app_ut_xi_u_sq_exp} into Eq.~\eqref{eq:app_trace_identity_proj} yields
\begin{equation}
  \mathbb E\,\mathrm{Tr}\!\big((P\xi P)^2\big)
  =(N-1)\left(N-2+\frac{2}{N}\right) \, ,
  \label{eq:app_proj_trace_result}
\end{equation}
and therefore
\begin{equation}
  R
  =2\sigma\sqrt{N-2+\frac{2}{N}} \, .
  \label{eq:app_R_dark}
\end{equation}
Notice that this expression is consistent with the one derived in Eq.~\eqref{eq:app_R_full} after sending $N \to N-1$, confirming that $P\xi P$ is still well described by a GOE matrix with vanishing diagonal. Indeed, the subleading $\frac{2}{N}$ correction, which we have dropped from Eq.~\eqref{eq:rho_general}, can be seen as the residual contribution coming from the projection onto the grey subspace.

From this expression, we can derive the fermionic polarization $k$ as in Eq.~\eqref{eq:k_strong_mu_sigma}, which without approximations reads
\begin{equation}
  k = (N-1) \left( \frac{\mu}{2}
  +\frac{4\sigma}{3\pi}\sqrt{N-2+\frac{2}{N}} \right).
  \label{eq:app_k_interpolation_refined}
\end{equation}
The corresponding condensate amplitude and boson occupation follow from
Eq.~\eqref{eq:app_nb_from_k}. In terms of the strong-coupling estimate
\eqref{eq:app_k_interpolation_refined}, they read
\begin{equation}
  |\alpha|
  =\frac{\sqrt{2}(N-1)}{\omega_0\sqrt{N}}
  \left(
    \frac{\mu}{2}
    +\frac{4\sigma}{3\pi}\sqrt{N-2+\frac{2}{N}}
  \right),
\end{equation}
and
\begin{equation}
  n_b
  =\frac{2(N-1)^2}{N\omega_0^2}
  \left(
    \frac{\mu}{2}
    +\frac{4\sigma}{3\pi}\sqrt{N-2+\frac{2}{N}}
  \right)^2.
\end{equation}
These expressions make explicit how the mixed strong-coupling regime
interpolates between the clean scaling $n_b\propto N$ and the
disorder-dominated scaling $n_b\propto N^2$.

\end{document}